\documentclass[preprint,tightenlines,aps,nofootinbib,showpacs]{revtex4}

\usepackage{graphicx}
\usepackage{dcolumn}
\usepackage{bm}
\usepackage{subfigure}
\usepackage{mathrsfs}
\usepackage{amssymb}
\usepackage{caption}
\usepackage{ragged2e}
\usepackage{multirow}
\usepackage{hyperref} 
\usepackage{color}
\usepackage[dvipsnames]{xcolor}
\usepackage[utf8]{inputenc}

\newcommand{\mev}{\ensuremath{\,\mathrm{MeV}}}
\newcommand{\gev}{\ensuremath{\,\mathrm{GeV}}}

\def\be{\begin{eqnarray}}
\def\en{\end{eqnarray}}
\def\non{\nonumber}
\def\la{\langle}
\def\ra{\rangle}
\def\bra{\langle}
\def\ket{\rangle}

\def\tcb#1{\textcolor{blue}{#1}}
\def\tcr#1{\textcolor{red}{#1}}
\def\tcg#1{\textcolor{green}{#1}}

\DeclareUnicodeCharacter{2212}{-}

\preprint{\font\fortssbx=cmssbx10 scaled \magstep2
\hbox to \hsize{
\hskip1.2in 
\hbox{\fortssbx \hspace{1.2in} }
\hskip0.2in $\vcenter{
                      \hbox{\bf arXiv: [hep-ph]}
                      \hbox{\bf OU-HEP-201228}
                      \hbox{December 2020}}$ }
}

\begin{document}

\title{\vspace*{0.7in}
Self-interacting Dark Matter with Scalar Dilepton Mediator}

\bigskip

\author{
  Chung Kao$^{a}$\footnote{E-mail address: Chung.Kao@ou.edu},
  Yue-Lin Sming Tsai$^{b,c}$\footnote{E-mail address: smingtsai@pmo.ac.cn},
  Gwo-Guang Wong$^{d}$\footnote{E-mail address: ggwong@hep1.phys.ntu.edu.tw}
}

\bigskip

\affiliation{
  $^a$Homer L. Dodge Department of Physics, University of Oklahoma,
  Norman, OK 73019, USA \\
  $^b$Key Laboratory of Dark Matter and Space Astronomy,
  Purple Mountain Observatory, Chinese Academy of Sciences,
  Nanjing 210033, China \\
  $^c$Department of Physics, National Tsing Hua University,
  Hsinchu 300, Taiwan \\
  $^d$Department of Physics, National Taiwan University, Taipei 10617, Taiwan
}
  
\bigskip


\bigskip

\begin{abstract}
The cold dark matter (CDM) candidate with weakly interacting massive
particles can successfully explain the observed dark matter relic
density in cosmic scale and the large-scale structure of the Universe.
However, a number of observations at the satellite galaxy scale seem
to be inconsistent with CDM simulation.
This is known as the small-scale problem of CDM.
In recent years, it has been demonstrated that
self-interacting dark matter (SIDM) with a light mediator offers
a reasonable explanation for the small-scale problem.
We adopt a simple model with SIDM and focus on the effects of
Sommerfeld enhancement.
In this model, the dark matter candidate is a leptonic scalar particle
with a light mediator. 
We have found favored regions of the parameter space with proper masses and
coupling strength generating a relic density that is 
consistent with the observed CDM relic density.
Furthermore, this model satisfies the constraints of recent direct searches
and indirect detection for dark matter 
as well as the effective number of neutrinos and the
observed small-scale structure of the Universe.
In addition, this model with the favored parameters can resolve the
discrepancies between astrophysical observations and $N$-body simulations.
\end{abstract}




\maketitle

\section{Introduction}

The first evidence of dark matter (DM) was observed by Fritz
Zwicky~\cite{Zwicky} in 1933.
The existence of dark matter can be observed in the whole
Universe, at the small galactic scale~\cite{RF,BBS},
the large scale of galaxy clusters~\cite{Carroll,CBGM}, and 
the cosmological scale~\cite{WMAPa,SDSS}.
The evidence of dark matter is usually inferred from its gravitational
interactions. However, weakly interacting massive particles (WIMPs)
provide intuitive candidates as cold dark matter (CDM). 
Stable invisible WIMPs with proper mass and coupling strength can lead
to a matter density  that is consistent with the observed DM relic
density in cosmic scale structure of the Universe~\cite{Aghanim:2018eyx}.
In addition, CDM can account for the consistency of large
scale structure ($\gtrsim$ 1Mpc) in the Universe between
the astrophysical observations~\cite{Colless:2001gk} and
$N$-body simulations~\cite{Springel:2006vs}.

There exist some discrepancies between CDM $N$-body simulations and
astrophysical observations on small scale structure of the Universe.
The first one is the cusp-core problem
(CCP)~\cite{Moore:1994yx,Salucci:2018hqu}.
The observed mass distributions are more flat in the central region
of dwarf galaxies without a steep cusp predicted from CDM
simulations.
The second one is the missing satellite problem (MSP)~\cite{Moore:1999nt}.
The observed number of dwarf satellite galaxies in the Milky Way (MW)
is much less than that predicted from CDM simulations. In recent year
there is another problem originally from the MSP, which is called the
too-big-to-fail (TBTF)~\cite{BoylanKolchin:2011de,BoylanKolchin:2011dk}.
Most massive sub-halos generated from the CDM $N$-body simulation are
too massive in the Milky Way halo with circular velocity larger than
30 km/s, whereas the observed maximum circular velocities of dwarf
spheroidals are less than 25 km/s.

All three problems, CCP, MSP and TBTF, are called the small
scale problem and they can be resolved if the CDM
particles are self-interacting with a light mediator to give a large
self-interacting cross section (SICS)~\cite{Spergel:1999mh}.
The large SICS provides a positive gradient of velocity
dispersion /temperature near the center of (sub)halo such that the heat
flow moves outward avoiding the formation of a density cusp until
self-interaction becomes weak and forms a flat core density.
Also, the smaller (sub)halo has a lower temperature since the faster DM
particles are easily to escape from the long-range gravitational
potential.
Hence, with the smaller subhalo in the host halo, the large SICS can
transfer heat from hotter DM particles in host halo to the colder DM
particle in subhalo resulting in the subhalo’s fragmentation or
evaporation.
On the other hand, the self-interacting cross section can not be
too large.
Otherwise, the small structures of our Universe such as
satellite galaxies would likely be washed out.  

From the astrophysical observation of galaxies and clusters of galaxy, we
have the following constraints.
\begin{itemize}
\item For a galaxy cluster with circular velocity $v\simeq$ 1000 km/s,
  the momentum transfer cross section ($\sigma_T$) per unit DM mass
  is $\sigma_T/m_\chi\simeq 0.1 \rm{\ cm^2/g}$~\cite{Rocha:2012jg}.
\item For a galaxy with $v\simeq$ 100 km/s, $\sigma_T/m_\chi\simeq 2
  \rm{\ cm^2/g}$.
\item For a dwarf galaxy with $v\simeq 10-100\ \rm{km/s}$,
  $\sigma_T/m_\chi\simeq 5-10 \rm{\ cm^2/g}$~\cite{Elbert:2014bma}.
\end{itemize}
We can see that the SICS increases with
decreasing DM velocity.
The velocity-dependent SICS is required to solve the problems occurred
in small structure scale of Universe.
In addition, possible anomalies in cosmic-ray, positron excess observed
by PAMELA~\cite{Adriani:2008zr}, AMS02~\cite{Adriani:2013uda},
ATIC~\cite{Chang:2008aa} and FermiLAT~\cite{Abdo:2009} can also be
explained by requiring that the present DM annihilation cross section
be 2$\sim$3 orders of magnitude greater than that in the freeze-out
stage~\cite{Feng:2009hw}.
This can be achieved by considering the Sommerfeld effect on a SIDM
with a light mediator and can also be constrained by these anomalous
observations of cosmic-ray~\cite{Kaplinghat:2015gha}.

A simple and elegant model with a self-interacting leptonic scalar
dark matter ($\chi$) and a light mediator ($\zeta$) was recently
proposed~\cite{Ma:2018pft} to provide a CDM candidate and to solve
the small scale problems of the Universe.
The light mediator ($\zeta$) could have a large production cross section
through s-wave Sommerfeld enhancement at late times.
If it decays to electrons and photons, it would change 
the history of gas ionization in the Universe,
disrupt the cosmic microwave background (CMB), 
and be ruled out by the precise cosmological data
now available~\cite{Bringmann:2016din}.
To satisfy cosmological requirements, the mediator ($\zeta$) is chosen
to have special Yukawa couplings such that it would not decay into
electrons and photons~\cite{Ma:2018pft}.
In our analysis, we have adopted this model with a focus on Sommerfeld
enhancement to determine the CDM relic density more precisely.
In addition, we find the allowed parameter space
that satisfies all constraints of (a) recent direct searches
(b) indirect detection experiments, (c) the observed relic density,
(d) effective number of neutrinos, and (e)
the astrophysical observation of small-scale structure of Universe.

This paper is organized as follows.
In Sec. II, we introduce a leptonic scalar dark matter (LSDM) model 
proposed by E. Ma~\cite{Ma:2018pft}.
Sec. III shows direct search results for leptonic scalar dark matter, the
spin-independent cross section of DM-nucleon elastic scattering
in the LSDM model, and compare it with XENON1T data~\cite{Xenon1T2018-SI}.
In Sec. IV, we present DM relic density as well as discovery potential
of indirect search for CDM.
We evaluate the relic density and effects of the Sommerfeld
enhancement in the LSDM model to compare with the observed DM
relic density~\cite{Tanabashi:2018oca,Aghanim:2018eyx}.
In addition, we compare our results of indirect search with
Fermi-LAT~\cite{Fermi-LAT2015,Fermi-LAT:2016uux} and
H.E.S.S.~\cite{HESS2016} astrophysical observations.
The cosmological constraints on right-handed neutrino and the small
scale requirements (CCP, MSP, TBTF) are discussed in Sec. V and Sec. VI,
respectively.
We show favored regions of parameter space in Sec. VII.
Conclusions are drawn in Sec. VIII.
Some useful formulae for Sommerfeld enhancement are presented in Appendix A.

\section{Leptonic Scalar Dark Matter Model}

  Recently, a simple and elegant model with a self-interacting
leptonic scalar dark matter (LSDM) was proposed~\cite{Ma:2018pft}.
This model is a simple extension of the Standard Model (SM)
with conservation of a $U(1)_{\rm L}$ lepton number.
There exist a singlet scalar ($\chi$) chosen to be the DM candidate
with $L=1$, one light singlet scalar ($\zeta$) as a mediator with $L=2$,
and three right-handed neutrinos: $\nu_{Ri}\ (i=1,2,3)$.
The lepton number conservation assures the stability of $\chi$ and no vacuum
expectation value developed from $\chi$ and $\zeta$ scalar fields.

The general scalar potential consisting of
$\chi$, $\zeta$ and the SM Higgs doublet is given by~\cite{Ma:2018pft} 
\be
V&=&\mu_0^2\Phi^\dagger\Phi+\mu_1^2\chi^*\chi+\mu_2^2\zeta^*\zeta+(\mu_{12}\zeta^*\chi^2+H.c.)
\non\\
&+&\frac{1}{2}\lambda_0(\Phi^\dagger\Phi)^2
+\frac{1}{2}\lambda_1(\chi^*\chi)^2
+\frac{1}{2}\lambda_2(\zeta^*\zeta)^2
\non\\
&+&\lambda_{01}(\Phi^\dagger\Phi)(\chi^*\chi)
+\lambda_{02}(\Phi^\dagger\Phi)(\zeta^*\zeta)
+\lambda_{12}(\chi^*\chi)(\zeta^*\zeta) \, .
\label{eq:potential}
\en
The scalar masses have the following relations 
\be
m_H^2 = \lambda_0{\rm v}^2 \simeq (125\ {\rm GeV})^2 \, , \qquad 
m_\chi^2 = \mu_1^2+\frac{1}{2}\lambda_{01}{\rm v}^2 \, , \qquad
m_\zeta^2 = \mu_2^2+\frac{1}{2}\lambda_{02}{\rm v}^2 , ,
\en
and the Higgs vacuum expectation value is ${\rm v} \simeq 246$ GeV.

  For simplicity, let us consider a CP-conserving scalar potential
with eight free real parameters:
\be 
m_\chi, m_\zeta, \mu_{12}, \lambda_{1}, \lambda_{2}, \lambda_{01}, \lambda_{02},~{\rm and}~\lambda_{12}.
\label{eq:inputs}
\en 
The values of $\mu_0$ and $\lambda_0$ are fixed by the minimization
condition of the scalar potential and the measured Higgs mass.
The $\mu_{12}\zeta^*\chi^2$ term serves as the source to enhance the
self-interaction of $\chi\chi^* \to \chi\chi^*$ through the exchange
of $\zeta$.
That leads to the dominant $t$-channel cross section
\be 
\sigma(\chi\chi^* \to \chi\chi^* )
 = \frac{\mu_{12}^4}{4\pi m_\zeta^4 m_\chi^2} \, ,
\en
where $\zeta$ is the light mediator~\cite{Ma:2018pft}.

\begin{figure}[b!]
\centering
\includegraphics[width=0.7\textwidth]{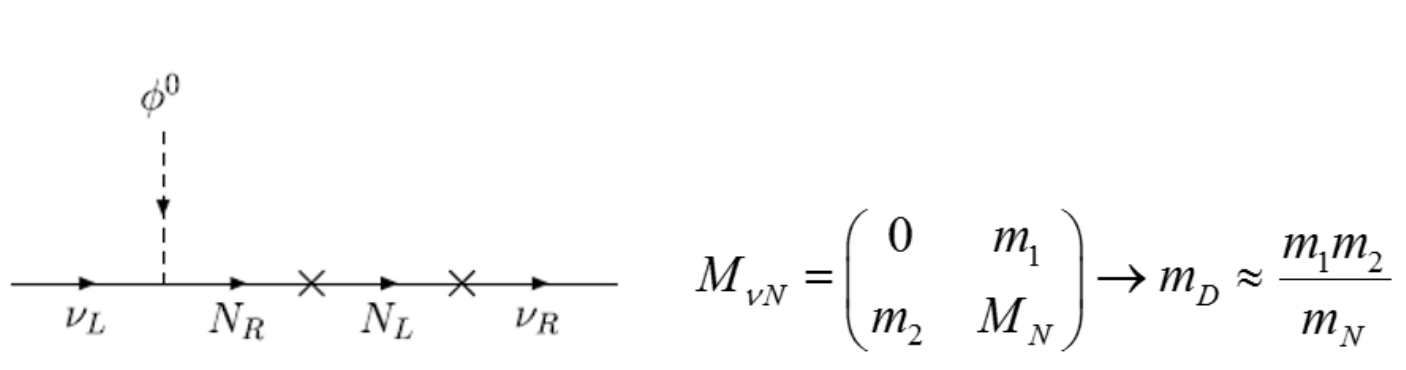}
\caption{Dirac neutrino mass with a Dirac singlet fermion insertion.} 
\label{fig: seesaw}
\end{figure}

The neutrinos in this model are Dirac fermions with small masses that 
could be natural consequences of various known
mechanisms~\cite{Bonilla:2016diq,Ma:2016mwh,Ma:2017kgb}.
For example, let us consider a discrete symmetry $S$,
such that under a transformation we have 
\be
\nu_L \to +\nu_L \, , \quad 
\phi^0 \to +\phi^0 \, , \quad \, , {\rm and} \quad 
\nu_R \to -\nu_R \, ,
\en
Then we insert a heavy singlet Dirac fermion $N$ with a large mass
$M_N$ as shown in Fig.~\ref{fig: seesaw}.
The S symmetry is softly broken by the dimension three
mass term $\bar{\nu}_R N_L$ with $N \to +N$ and $\nu_R \to -\nu_R$.    
The small masses $m_1$ and $m_2$ are generated by electroweak symmetry
breaking or soft $S$ symmetry breaking. 
That leads to a small Dirac neutrino mass through the see-saw mechanism.
Thus, the only new Yukawa couplings are 
\be
{\cal L}_Y = f_{ij}\zeta^*\overline{\nu_R^{\ c}}\nu_R +H.c. \, .
\en
After $\chi$ freezes out, $\zeta$ eventually decays to neutrinos
via the $f_{ij}$ terms with a lifetime
\be
\tau_\zeta = \frac{4\pi}{m_\zeta f^2_{RR}},
\en
where $f^2_{RR} \equiv \sum_{i,j}|f_{ij}|^2$.
We can see that $\tau_\zeta\lesssim 10^{-11}$ sec, for
$m_\zeta > m_\zeta^{\rm min} = 0.2$ GeV and
$f_{RR} > f_{RR}^{\rm min} = 10^{-6}$.
This means that before the onset of big bang nucleosynthesis (BBN),
all $\zeta$'s decay away and $\nu_R$ decouples from the SM particles
at the temperature $T_f^R \simeq m_\zeta$.

Before going further, we should note that $\chi^*$ has a lepton number
$L = -1$ and it is the antiparticle of $\chi$ with $L = +1$.
Since CP is conserved, the transition amplitude ($M_{\chi^*}$)
involving $\chi^*$ is the same as the transition amplitude
($M_{\chi}$) with $\chi$.
Now let $n_{\chi}$ and $n_{\chi^*}$ be the number density of the particle
$\chi$ and the antiparticle $\chi^*$, respectively. Their sum is
total number density of DM, $n=n_{\chi}+n_{\chi^*}$.
Both $\chi$ and $\chi^*$ make equal amount of contributions to the
annihilation cross section and the CDM relic density.

\section{Direct Search for Leptonic Scalar Dark Matter}

  It is an opportune time to investigate direct detection for WIMP
dark matter. The XENON1T collaboration recently announced the
observation of low energy excess electronic recoil events
in their detector~\cite{Aprile:2020tmw}.
In this section, we focus on the search for nuclear recoils generated by
the WIMP-nucleon scattering.
We evaluate $\chi$-nucleon elastic scattering cross section
for the leptonic scalar dark matter ($\chi$).
In addition, we place limits on the relevant parameters $\lambda_{01}$
and $m_\chi$ with XENON1T results.

\subsection{The Elastic Scattering of DM with Nucleus ${\cal N}$}

In the LSDM model, the leptonic scalar DM ($\chi$) interacts with
quarks ($q$) through the SM Higgs boson ($H$).
Hence the effective Lagrangian of $\chi$ with quarks is simply
\be
{\cal L}_{\rm eff} =
\sum_{q} a_q[\chi\chi^*][{\bar q}q] \, , \qquad {\rm with} \quad 
  a_q \simeq \lambda_{01} \frac{m_q}{m_H^2}  \, ,
\label{eq:Leff}
\en
where $m_q$ and $m_H$ are masses of the quark and
the Higgs boson, respectively, and $a_q$ is the effective 
coupling of DM interacting with the quark in a nucleon.

Before making an estimate for the detection rate of the leptonic
dark matter ($\chi$) in the XENON1T experiment, let us evaluate the
normalized spin independent (SI) cross section for the leptonic scalar DM
scattered off the $^{129,131}{\rm Xe}$ nuclei.
Applying the effective Lagrangian in Eq.~(\ref{eq:Leff}), we obtain
the matrix element for elastic scattering of the DM particle ($\chi$)
and the nucleus (${\cal N}$) as
\be
M_{fi} =
 2m_{\cal N}\sum_q a_q\langle {\cal N}_f|{\bar q}q|{\cal N}\rangle \, .
\en
In general, the averaged unpolarized amplitude square at $q^2=0$
can be written as 
\be
\overline{\sum}|M_{fi}|^2(q^2=0)
& = & \overline{\sum}|M^{SI}_{fi}|^2(q^2=0)
     +\overline{\sum}|M^{SD}_{fi}|^2(q^2=0) \non \\
& = & 4m_{\cal N}^2 f^2_{s\cal N} \, ,
\label{eq:iMsq}
\en
where $SI$ and $SD$ denote the spin independent and the spin dependent
contributions, respectively.


For the effective scalar interaction, we have
\be
  f_{s\cal N} = Z f_{sp}+(A-Z) f_{sn},
\label{eq: fsN2}
\en
where
\be
f_{s p(n)} = \sum_{q=u,d,s} a_q\frac{m_{p(n)}}{m_q} f^{(p(n))}_{Tq}+
 \sum_{q=c,b,t} a_q\frac{2}{27} \frac{m_{p(n)}}{m_q}
 \left(1-\sum_{q'=u,d,s} f^{p(n)}_{Tq'}\right) \, .
\label{eq:MAq=0}
\en
When evaluating the quark operator matrix element in the nuclear
state, we need to include loop contributions involving
heavy quarks that contribute to the mass of the nucleon ($m_{p(n)}$).

The proton mass fraction $f^{p}_{Tq}$ is defined by the matrix
elements of the quark current
\be 
\la p|m_q {\bar q}q|p\ra=\left\{
\begin{array}{lr}
m_p f^{p}_{Tq},
& q=u,d,s\, , \\
\frac{2}{27} m_p \left(1-\sum_{q=u,d,s} f^{p}_{Tq}\right),
&q=c,b,t \, .
\end{array}
\right.
\en
The neutron mass fraction $f^{n}_{Tq}$ is defined in the same way.
The matrix elements of the light-quark currents in the proton or
neutron are obtained in chiral perturbation theory from measurements
of the pion-nucleon sigma
term~\cite{Cheng1,Cheng2,GLS,Alarcon2011,Alarcon2012,Cheng3}.
The heavy quark contribution to the mass of the nucleon is through
the triangle diagram~\cite{SVZ}.

In the center of mass (CM) frame, the differential cross section is 
\be 
\frac{d\sigma(\vec q=0)}{d|{\bf q}|^2}
= \frac{1}{64\pi s\mu^2_{\cal N}v^2}\overline{\sum}|M_{fi}|^2(q^2=0)
\, ,
\en
where $v$ is the DM velocity relative to the target,
$\sqrt{s}\approx m_\chi+m_{\cal N}$ is the total energy,
and $\mu_{\cal N}$ is the reduced mass of DM and the target nucleus $\cal N$. 
The total cross section at zero momentum transfer~\cite{JKG}
can then be obtained as 
\be 
\sigma_0^{SI}
= \int^{4 \mu^2_{\cal N}v^2}_0d|{\bf q}|^2
  \frac{d\sigma(\vec q=0)}{d|{\bf q}|^2}
=\frac{\mu_{\cal N}^2}{\pi} f^2_{s\cal N} \, .
\label{eq: sigmaSISD}
\en
Hence the total cross section of DM-nucleus ($\chi-{\cal N}$)
scattering becomes
\be 
\sigma_{{\cal N}}=\frac{\sigma^{SI}_0}{4\mu^2_{{\cal N}} v^2}
\int^{4\mu^2_{{\cal N}}v^2}_0d|{\bf q}|^2 F^2_{SI}(|{\bf q}|)
\label{eq: sigmaN}
\en
where $F^2_{SI}(|{\bf q}|)$ is the spin-independent form factor.  
To compare with the experimental results, we define the scaled SI and
SD cross sections, respectively, for the nucleus with atomic mass
number $A_i$ and isotope abundance $\eta_i$ as the following
\be
\sigma^{SI}_{\chi p}&\equiv&\frac{\sum_i \eta_i\sigma_{A_i}}
{\sum_j \eta_j  A^2_j\frac{\mu^2_{A_j}}{\mu^2_p}},
\label{eq: sigma Z N}
\en
and 
\be
\sigma^{SD}_{\chi p,n}\equiv (\sum_i\eta_i\sigma_{A_i})
\left(
\sum_j \eta_j\frac{4\mu_{A_j}^2 \la S_{p,n}\ra^2_{\rm eff}  (J_{A_j}+1)}{3\mu_{p,n}^2J_{A_j}}\right)^{-1},
\label{eq: SD p n}
\en
where $\mu_{A_i}$ and $\mu_{p,n}$ are the reduced masses of the DM
with the target nucleus and the DM with proton or neutron, respectively.
In the above, $\la S_{p(n)}\ra_{\rm eff}$ and ($J_{A_j}$) are the
proton (neutron) spin expectation value (including the contributions
of two-body current~\cite{Menendez}) and the total angular momentum of
the nucleus with atomic mass number $A_j$ respectively.
The effective spin expectation value is defined as
$\la S_{p(n)}\ra_{\rm eff}\equiv \la S_{p(n)}\ra \pm \delta a_1(\la
S_p\ra -\la S_n\ra)/2$ and $\delta a_1$ is the fraction contributing
to the isovector coupling~\cite{ChuaWong,Menendez}.

\subsection{Numerical Results for Direct Search}

At present, the XENON1T experiment~\cite{Xenon1T2018-SI} provides
the most stringent upper limits on $\sigma^{SI}$ for WIMP masses above 6 GeV.
In our analysis for spin independent cross section of $\chi-{\cal N}$
scattering, we adopt the Helm form factor~\cite{LS,VKMHS} used
in XENON1T experiments:
\be 
F^2_{SI}(|{\bf q}|)
 = \bigg( \frac{3j_1(qR_{\cal N})}{qR_{\cal N}}\bigg)^2e^{(qs)^2} \, ,
\en
where the nuclear radius $R_{\cal N}=c^2+\frac{7}{3}\pi^2a^2-5s^2$
with $c=(1.23A^{1/3}-0.6)$ fm, $a=0.52$ fm and
the nuclear surface thickness $s=1$ fm. 
We use the updated data of nucleon mass fractions from Ref.~\cite{Cheng3}: 
$f^p_{Tu} = 0.017$,
$f^p_{Td} = 0.023$,
$f^n_{Tu} = 0.012$,
$f^n_{Td} = 0.033$,
$f^{p,n}_{Ts} = 0.053$.

In the LSDM model with a scalar dark matter ($\chi$)
and a light mediator ($\zeta$), there are eight free parameters
as shown in Eq.~(\ref{eq:inputs}).
In our analysis, the scan is performed with the log-prior distributions 
for the input parameters as shown in the below:
\begin{itemize}
\item $m_H/2 \le m_\chi \le 1$ TeV, such that $\chi\chi\to\zeta H$ can
  occur,
\item $0.2$ GeV $\le m_\zeta \le 1.2$ GeV,
\item $1$ GeV $\le \mu_{12} \le 1$ TeV, and 
\item $10^{-6} \le \lambda \le {\cal O}(1) \sim \sqrt{4\pi}$ for
      $\lambda = \lambda_1, \lambda_2, \lambda_{01}$, or
  $\lambda_{12}$. 
\end{itemize}
Note that $\lambda_{02}$ is chosen to be
$10^{-6} \le \lambda_{02} \le 10^{-2}$.
It is constrained by the SM Higgs invisible
decay width ($H\rightarrow \zeta\zeta^*$), i.e.
\be
  \Gamma(H\rightarrow \zeta\zeta^*)
  = \frac{\lambda_{02}^2 v^2}{16\pi m_H} \, .
\en
Assuming that the invisible width is less than $10\%$ of
the Higgs width $\Gamma_H \sim 4.12$~MeV~\cite{Tanabashi:2018oca},
we obtain the maximal value of $\lambda_{02} \sim 6.5\times 10^{-3}$.
In addition, $m_\zeta$ must be greater than 0.2 GeV to satisfy the
cosmological constraint of effective number of neutrinos,
which will be discussed later.

\begin{figure}[h!]
\centering
\captionsetup{justification=raggedright}
\subfigure[]
{
 \includegraphics[width=0.45\textwidth]{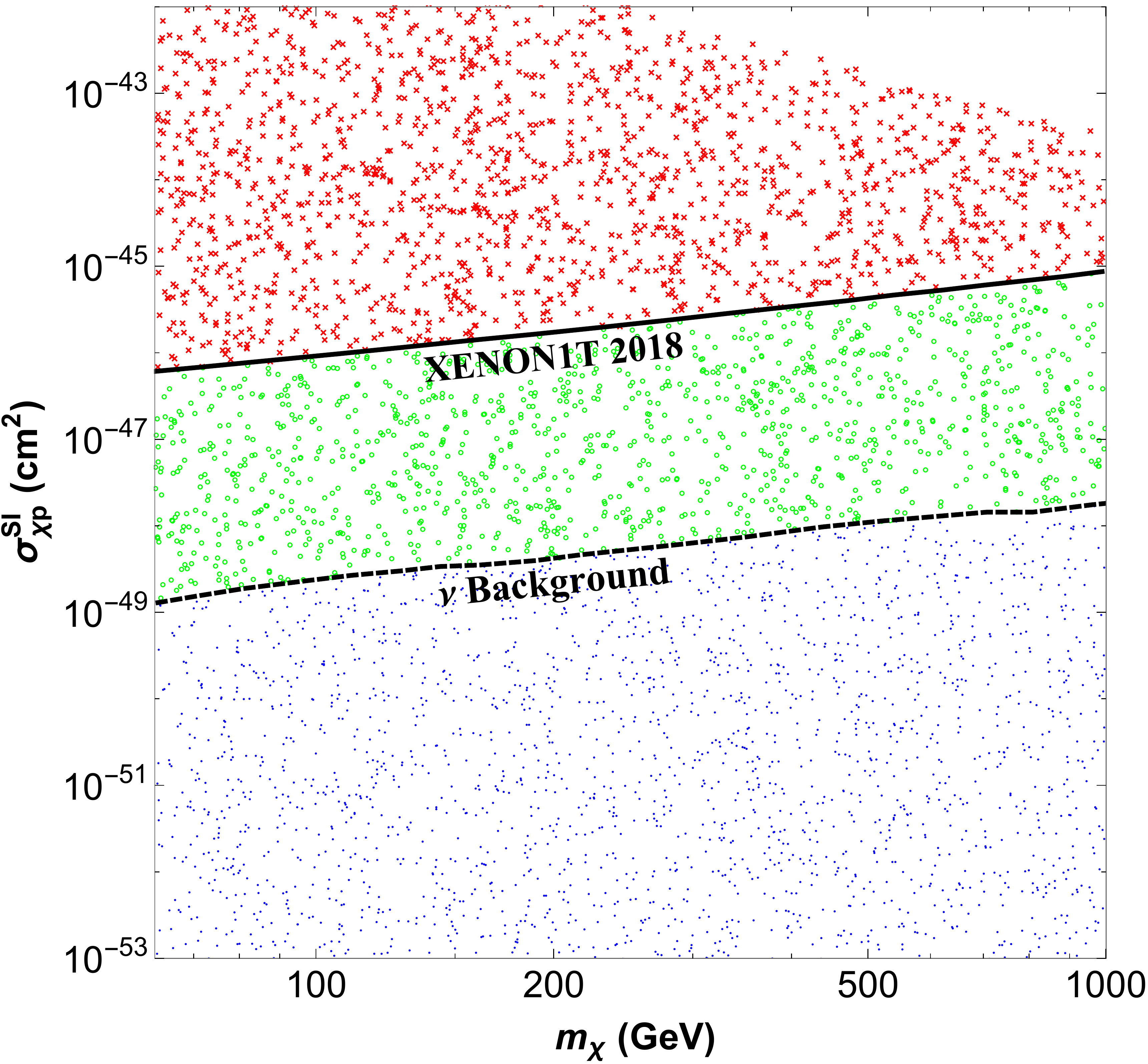}
}
\subfigure[]{
  \includegraphics[width=0.45\textwidth]  {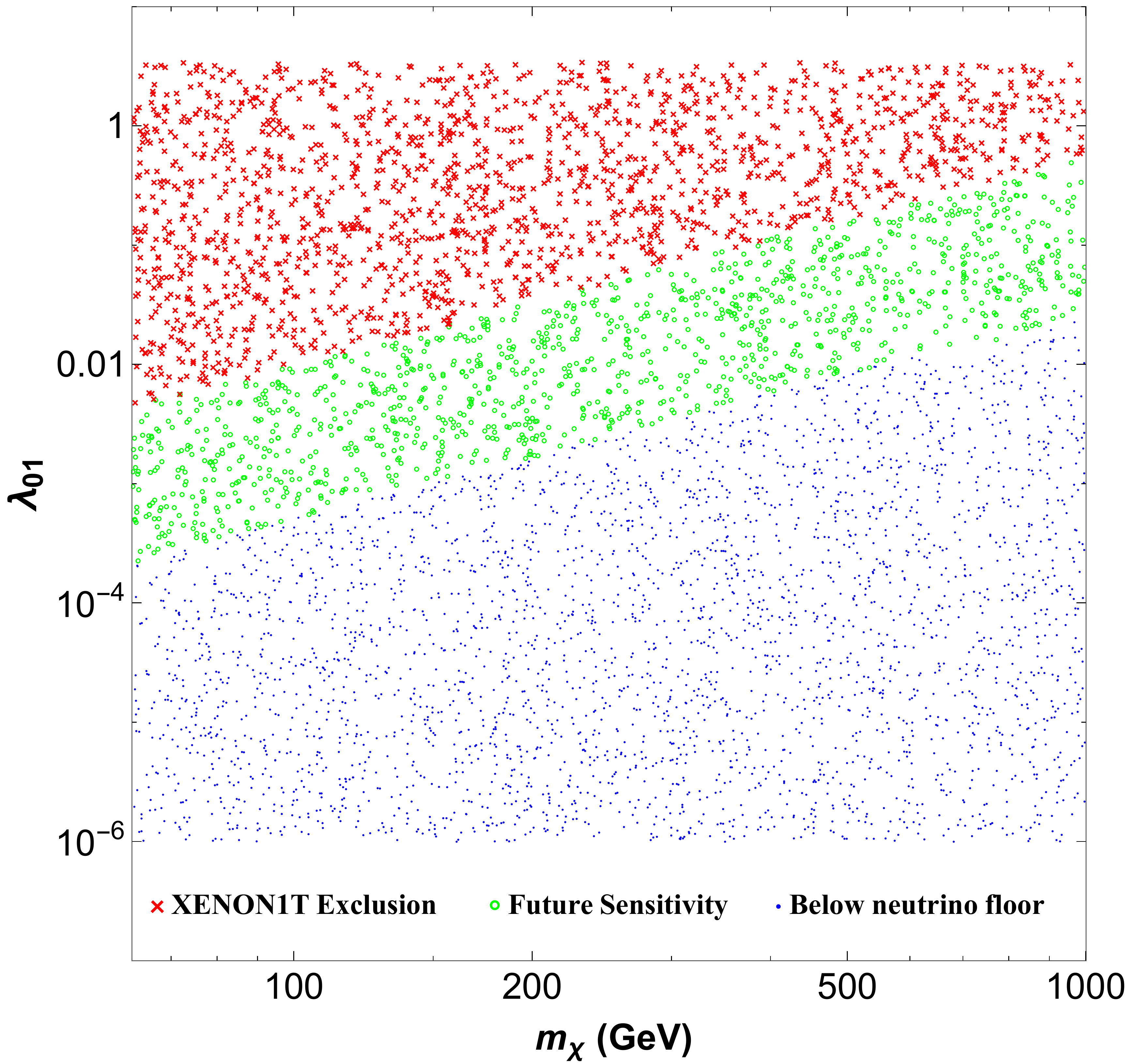}
}
\caption{
  (a) The spin independent cross section $\sigma^{SI}_{\chi p}$
  versus $m_\chi$ with random sampling for DM ($\chi$) scattered off
  the nuclei $^{129,131}$Xe.
  Also shown are the upper limit from XENON1T~\cite{Xenon1T2018-SI}
  and the neutrino background~\cite{nb3}.
  (b) A scatter plot for the same samples projected to the
  the plane of ($m_\chi, \lambda_{01}$) with the corresponding
  $\lambda_{01}$.}
\label{fig: direct}
\end{figure}

In this model, the scaled SI and SD cross sections 
($\sigma^{SI}_{\chi p}$ and $\sigma^{SD}_{\chi p,n}$) depend on
two parameters: (i) the mass of leptonic scalar dark matter ($m_\chi$), and
(ii) the coupling $\lambda_{01}$.
We apply the most stringent constraint from XENON1T
experiment~\cite{Xenon1T2018-SI} with the upper limits of SI
WIMP-nucleon scattering cross section.
Fig.~(\ref{fig: direct}a) shows the spin independent cross section
$\sigma^{SI}_{\chi p}$ versus $m_\chi$ with random sampling
for DM ($\chi$) scattered off the nuclei $^{129,131}$Xe.
In addition, a scatter plot for the same samples projected to the
the plane of ($m_\chi, \lambda_{01}$) with the corresponding
$\lambda_{01}$ is presented in Fig.~(\ref{fig: direct}b).
In this figure, we show  three groups of samples:
(a) all samples with red ``$\tcr{\times}$" above the the upper
limits of  XENON1T experiment~\cite{Xenon1T2018-SI} are ruled out,
(b) those with green ``$\tcg{\circ}$" between the upper limits of XENON1T
experiment and the curve of neutrino background~\cite{nb1,nb2,nb3}
are allowed and could be detectable in future detectors, and
(c)  the samples with blue ``$\tcb{\cdot}$" below
the curve of neutrino background, and they are allowed as well.
However, we may not be able to distinguish the DM event from neutrino
event.

\section{ Relic Density and Indirect Search }

  The matter density ($\rho$) of the Universe is often described with
  a relative density ($\Omega$)
\be
\Omega  & = & \frac{\rho}{\rho_{\rm c}} \\
\rho _c & = & \frac{3H_0^2}{8\pi G_N}
      \simeq 1.88 \times 10^{-29} h^2 \; {\rm g}/{\rm cm}^3
\en
where $\rho_{\rm c}$ is the critical density, $G_N$ is Newton's
gravitational constant, and $H_0$ is the Hubble
constant, conventionally expressed as 
\be
H_0 = 100 h~{\rm km}/s/{\rm Mpc} \, ,
\en
and $h \simeq 0.68$~\cite{Tanabashi:2018oca}.
   
The Planck collaboration has measured cosmological parameters with
very high precision~\cite{Aghanim:2018eyx}. The updated cold dark
matter relic density~\cite{Aghanim:2018eyx} is 
\be
\Omega_{\rm CDM} h^2 = 0.120 \pm 0.001 \, .
\en
We can also take a conservative approach as demonstration that 
$\chi$ can be produced again in the late time by other cosmological mechanisms 
so that the relic density at the present at $3\sigma$ allowed range follows
\be
\Omega_\chi h^2 \le 0.123 \, . 
\en
This assumption also includes the standard scenario $\Omega_\chi h^2 \approx 0.12$.

\subsection{Thermal Relic Dark Matter Density}

In the early Universe, DM $\chi$ existed abundantly in thermal
equilibrium with other particles. The evolution of the total number
density ($n(t) = n_{\chi}+n_{\chi^*} = 2n_{\chi}$) for the leptonic
dark matter is described by the Boltzmann equation:
\begin{equation}
  \frac{dn}{dt} + 3Hn
  = -\langle\sigma_\mathrm{ann} v\rangle
  [n^2 - n^2_\mathrm{E}] \, ,
\label{eq: Boltzmann}
\end{equation}
where $n_\mathrm{E}$ is the number density at thermal equilibrium,
the Hubble parameter is
\be
H = \sqrt{ 4\pi^3g_*(T)\, T^4/(45M_{\mathrm{Pl}}^2) } 
  \simeq 1.66 g_*^{1/2} T^2/M_{\mathrm{Pl}} \, ,
\en
$M_{\mathrm{Pl}}=1.2\times 10^{19}$ GeV is the Planck mass, 
$g_*$ is the total effective number of relativistic degrees of
 freedom~\cite{Kolb,CR:03},
$\la \sigma_\mathrm{ann} v \ra$ is the thermally averaged
annihilation cross section times velocity,
and $v$ is the relative velocity. 
The relative velocity 
\begin{equation}
v \equiv v_{\rm lab} = \sqrt{s(s-4m_{\chi}^2)}/(s-2m_{\chi}^2)
\label{eq:vlab}
\end{equation}
and the Mandelstam variable $s = 2m_{\chi}^2(1+1/\sqrt{1-v^2)}$ are
measured in the lab frame.

\begin{figure}[t!] \centering
\includegraphics[width=0.8\textwidth]{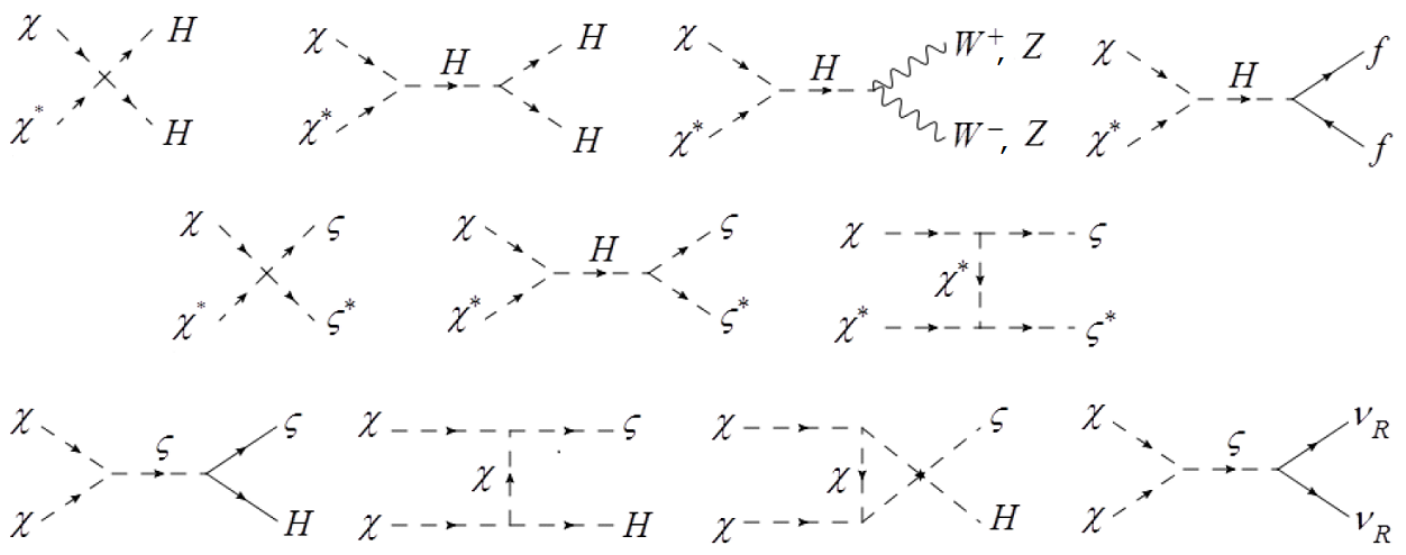}
\caption{Feynman diagrams for leptonic scalar DM annihilation.}
\label{fig: FeynDiagram}
\end{figure}

The thermally averaged annihilation cross section times velocity
$\bra \sigma_\mathrm{ann} v \ket$ is evaluated with 
the Maxwell velocity distribution,
\begin{eqnarray}\label{T_average}
\langle\sigma_{\mathrm{ann}}v\rangle
& = &\frac{3\sqrt 6}{\sqrt{\pi}v_0^3}\int_0^\infty dv\, v^2
\frac{(\sigma_{\mathrm{ann}}v)_{\chi\chi^*}}{2}
e^{-3v^2/2v_0^2}
\non\\
& = &
\frac{x_f^{3/2}}{2\sqrt{\pi}}\sum_{\varphi_1,\varphi_2}
\int_0^\infty dv\, v^2
\frac{[\sigma_{\mathrm{ann}}(\chi\chi^*\to
    \varphi_1\varphi_2)v]}{2}e^{-xv^2/4} \nonumber \\
&   & \quad \quad
      \times \theta[2m_\chi^2(1+\frac{1}{\sqrt{1-v^2}})-(m_{\varphi_1}+m_{\varphi_2})^2]
      \, ,
\label{eq: Jfactor}
\end{eqnarray}
where $x_{f}\equiv m_\chi/T_{f}$, $v_0\equiv\la v^2\ra^{1/2}=\sqrt{6/x_f}$
with the freeze-out temperature $T_f$. 
The second expression represents the leading contribution of DM
annihilating to a pair of particles ($\varphi_1$ and $\varphi_2$) in the final state.
Fig.~\ref{fig: FeynDiagram} shows the Feynman diagrams for the
dominant leptonic scalar DM annihilation processes.
For each annihilation channel, we have put a step function ($\theta$)
for the threshold energy.

As the Universe cooled down, deviation of number density ($n_\chi$)
from thermal equilibrium would begin when the temperature reached
the freeze-out temperature ($T_f$).
After the temperature dropped to approximately $T_f/5$,
the annihilation rate of the WIMP dark matter became equal to
the expansion rate of universe~\cite{JKG}, namely
$n_\chi \bra \sigma_\mathrm{ann} v \ket = H$.
The relic mass density becomes
\be
\Omega_\chi h^2 = n_\chi m_\chi/\rho_{\rm c}
 = \frac{H m_\chi}{\bra \sigma_\mathrm{ann} v \ket \rho_{\rm c}}
\en
and the leptonic scalar $\chi$ remains as cold dark matter. 
From the freeze-out condition, $n_\chi \bra \sigma_\mathrm{ann} v \ket = H$,
the freeze-out temperature parameter $x_f$ can be solved numerically
by the following equation~\cite{Kolb,JKG}
\be
x_f={\rm ln}\left [c(c+2)\sqrt{\frac{45}{8}}
  \frac{g_{\chi} m_{\chi} M_{\rm pl}(a+6b/x_f)}
  {2\pi^3\sqrt{g_*(m_{\chi}/ x_f)}x_f^{1/2}}\right ] \, .
\label{eq: xf}
\en
In the above, $c$ is an order of unity parameter determined by
matching the late-time and early-time in the freeze-out criterion. 
The exact value of $c$ is not so significant to solve the numerical
solution for $x_f$ due to the logarithmic dependence in Eq.~(\ref{eq: xf}). 
Following the standard procedure~\cite{Kolb} to solve
the Boltzmann equation [Eq.(\ref{eq: Boltzmann})], 
the relic DM density $\Omega_{\rm DM}\equiv \rho_\chi / \rho_{\rm  c}$
can be approximately related to 
$\la\sigma_{\rm ann}v\ra$ as
\begin{equation}\label{abundance}
\Omega _{\text{DM}}h^2 \approx 1.04\times
10^9\frac{{\rm GeV}^{-1}}{M_{\mathrm{Pl}} \sqrt{g_*\left(T_f\right)}J(x_f)},
\end{equation}
where
\begin{equation}\label{Jfactor}
J\left(x_f\right) 
\equiv \int_{x_f}^{\infty }
\frac{\langle\sigma_{\mathrm{ann}}v\rangle}{x^2} dx
=\int_0^\infty dv\frac{(\sigma_{\rm ann} v)_{\chi\chi^*}}{2} v\left[1-{\rm erf}\left(v\sqrt x_f/2\right)\right].
\end{equation} 
In the non-relativistic limit, $J(x_f)=a/x_f+3 b/x_f^2$.

The DM particles became non-relativistic (NR) when they froze out of
thermal equilibrium in the early universe.
In the NR limit, we have
$\sigma_\mathrm{ann}(\chi\chi^*\rightarrow {\rm all})v = a +b v^2 +O(v^4)$
by applying the Taylor series expansion 
and its thermally-averaged value Eq.~(\ref{eq: Jfactor}) can be simplified as 
$\la\sigma_\mathrm{ann}v\ra= a+6b/x_f+O(1/x_f^2)$.

\subsection{Indirect Search for Leptonic Scalar Dark Matter}

In the halo of the Milky Way and nearby galaxies, 
WIMP DM annihilation might generate high energy gamma-rays 
and appear in detectors such as
Fermi-LAT~\cite{Fermi-LAT2015,Fermi-LAT:2016uux}, 
H.E.S.S.~\cite{HESS2016,Rinchiuso:2019rrh},
HAWC~\cite{Albert:2017vtb},
MAGIC~\cite{Ahnen:2017pqx}, or
VERITAS~\cite{Archambault:2017wyh}.
In addition, WIMP dark matter
would lose energy when they pass through massive stars such as the sun.
They become gravitationally trapped and accumulate. WIMP annihilations
could be sources of high energy neutrinos and might be detected by
ANTARES~\cite{ANTARES:2019svn} and IceCube~\cite{Aartsen:2020tdl}.

At present, the most stringent limits for our surveyed DM mass range,
$m_H/2 < m_\chi < 1000$ GeV, come from 
Fermi-LAT~\cite{Fermi-LAT2015,Fermi-LAT:2016uux} and
H.E.S.S.~\cite{HESS2016}.
We will evaluate the leptonic scalar DM annihilation cross section $\langle\sigma_{\rm ann} v\rangle$ in different channels,
and investigate the discovery potential as well as determine
favored parameters guided by the Fermi-LAT and the H.E.S.S. data.

Fermi-LAT analyzed 15 dwarf spheroidal satellite galaxies
(dSphs)~\cite{Fermi-LAT2015,Fermi-LAT:2016uux},
while the H.E.S.S observed $\gamma$-ray towards the inner 300 parsecs
of the Milky Way.
The speed of the sun moving around the galactic center is
approximately 220 km/s at the local distance $r\approx$ 8.5 kpc and
the galactic circular rotation speed is about 230 km/s at radii $\approx$ 100
kpc~\cite{JKG,Kochanek:1995xv}. On the other hand,
the distance between the 15 dSphs and the sun is $\approx 23--233$
kpc~\cite{Fermi-LAT2015}.
In the indirect-detection calculation,
we conventionally adopt a typical DM velocity $v_0\simeq 10^{-3} $
in the unit of the light speed~\cite{Ferrer:2013cla}.

In the leptonic scalar DM model,
the dark matter particle ($\chi$) can annihilate into a pair of SM
particles such as 
$W^+W^-$, $Z^0Z^0$, $HH$, fermion pairs $f\bar f$, or $\zeta\zeta^*$ through
s-channel exchange of SM Higgs boson $H$. In addition, The leptonic
scalar DM can also annihilate into a pair of $\zeta\zeta^*$ or $HH$ through
4-point interactions and t-channel exchange of $\chi$,
or $\nu_R\nu_R$ through s-channel exchange of $\zeta$, or a pair of
$\zeta H$ through s-exchange of $\zeta$ and  t- and u-channel exchange
of $\chi$ as presented in Fig.~\ref{fig: FeynDiagram}.
From these Feynman diagrams, we calculate the corresponding DM
annihilation cross sections at tree-level:
\be
(\sigma v)_{\chi\chi^*\rightarrow\zeta\zeta^*}
& = & \frac{\sqrt{s-4m_\zeta^2}}{16\pi\sqrt{ s} (s-2m_\chi^2)}
\left\{ 
\lambda_{12}^2
+\frac{16\mu_{12}^2}{m_\zeta^4-m_\chi^2(s-4 m_\zeta^2)}\right. \nonumber \\
&   & \quad \quad +\frac{\lambda_{01}^2\lambda_{02}^2v^4
-2\lambda_{01}\lambda_{02}\lambda_{12}v^2(s-m_H^2)}
{(s-m_H^2)^2+m_H^2\Gamma_H^2} \non \\
&   & \quad \quad 
+\frac{8\mu_{12}^2\left[\lambda_{12}\left((s-mH^2)^2+m_H^2\Gamma_H^2)
-\lambda_{01}\lambda_{02}v^2(s-M_H^2\right)\right]}
{\sqrt{(s-4m_\chi^2)(s-4m_\zeta^2)}\left((s-mH^2)^2+m_H^2\Gamma_H^2\right)}
 \nonumber \\
&   &\left. \quad \quad
 \times \log\left[\frac{s-2m_\zeta^2+\sqrt{(s-4m_\chi^2)(s-4m_\zeta^2)}}
   {s-2m_\zeta^2-\sqrt{(s-4m_\chi^2)(s-4m_\zeta^2)}}\right]\right\},
\label{eq: sigma v1}
\en
\be
&&(\sigma v)_{\chi\chi^*\rightarrow f\bar f}
=\frac{\mathcal{C}_f\lambda_{01}^2m_f^2(s-4m_f^2)^{3/2}}
{8\pi\sqrt{s}(s-2m_\chi^2)
[(s-m_H^2)^2+m_H^2\Gamma_H^2]},\\ \non\\
&&
(\sigma v)_{\chi\chi^*\rightarrow VV}
=\frac{\lambda_{01}^2\sqrt{s-4m_V^2}(s^2-4 s m_V^2+12m_V^4)}
{16\pi\sqrt{s}(s-2m_\chi)[(s-m_H^2)^2+m_H^2\Gamma_H^2]}{S}
,\quad {S}=\left\{ \begin{array}{rcl}
1 & \mbox{for}
& W^+W^- \\ 1/2 & \mbox{for} & \ Z^0Z^0
\end{array}\right.\\ \non\\
&&
(\sigma v)_{\chi\chi\rightarrow \nu_{Ri}\nu_{Rj}}
=\frac{\mu_{12}^2|f_{ij}|^2\sqrt{s-4m_{\nu}^2}(s-2m_\nu^2)}
{\pi \sqrt{s}(s-2m_\chi^2)[(s-m_\zeta^2)^2+m_\zeta^2\Gamma_\zeta^2]},\\ \non\\
&&
(\sigma v)_{\chi\chi\rightarrow \zeta H}=
\frac{(\lambda_{01} \mu_{12} v)^2\sqrt{A}}{2\pi (s-2m_\chi^2)(m_\chi^2 A+s m_H^2m_\zeta^2)}
+\frac{2(\lambda_{01} \mu_{12} v)^2 s \tanh^{-1}(\sqrt{A B}/{C})}{\pi (s-2m_\chi^2)\sqrt{B}C}\non\\
&&\qquad\qquad\qquad -
\frac{2\lambda_{01}\lambda_{02}(\mu_{12} v)^2 (s-m_\zeta^2)\tanh^{-1}(\sqrt{A B}/{C})}{\pi (s-2m_\chi^2)\sqrt{B}D}
+\frac{(\lambda_{02} \mu_{12} v)^2\sqrt{A}}{4\pi s(s-2m_\chi^2)D},
\label{eq: sigma v5}
\en
where
\be
A&=&[s-(m_H+m_\zeta)^2][s-(m_H-m_\zeta)^2],\non\\
B&=&s(s-4m_\chi^2),\non\\
C&=&s(s-m_H^2-m_\zeta^2),\non\\
D&=&(s-m_\zeta^2)^2+m_\zeta^2\Gamma_\zeta^2.
\en
After substituting $s=2m_\chi^2(1+1/\sqrt{1-v^2})$ into the above
equations and expanding around $v^2$, one can obtain
the usual form:
$\langle\sigma_{\mathrm{ann}}v\rangle = \langle a+ b v^2 + {\cal
  O}(v^4)\rangle $ in the non-relativistic limit.
As discussed in Sec. II, the decay lifetime of $\zeta$ is so
short that the $\zeta$ has all decayed into two right-handed neutrinos $\nu_R$,
and the light $\nu_R$ decouples from the SM particles at its freeze out temperature 
$T^R_f\sim m_\zeta$ before the onset of BBN.

\subsection{Sommerfeld Enhancement Effect}

\begin{figure}[htb]
 \centering
 \captionsetup{justification=raggedright}
 \subfigure[]{
   \includegraphics[width=0.3\textwidth,height=0.12\textheight]{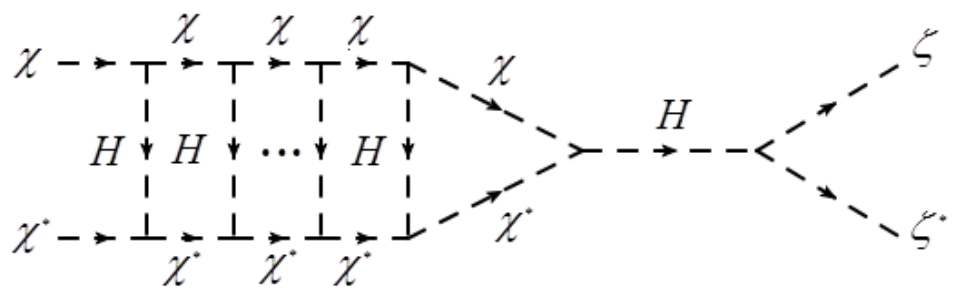}
 }
 \subfigure[]{
   \includegraphics[width=0.3\textwidth,height=0.12\textheight]{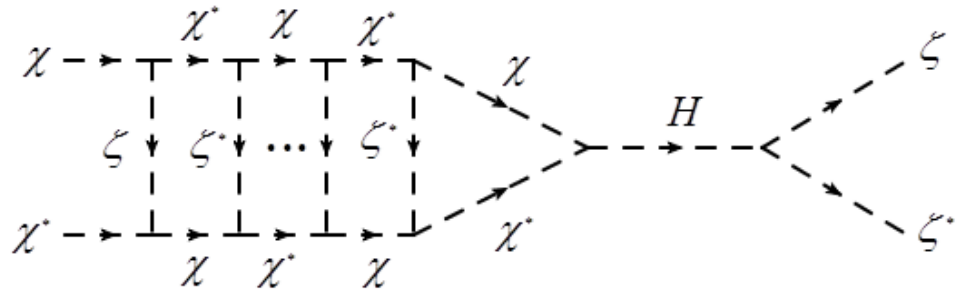}
 }
 \subfigure[]{
   \includegraphics[width=0.3\textwidth,height=0.12\textheight]{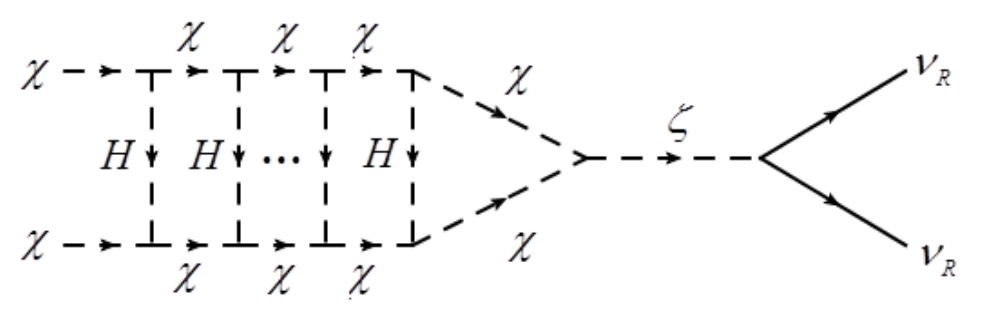}
 }
 \caption{Schematic diagrams for non-pertubative annihilation processes:
   $\chi\chi^*\rightarrow \zeta\zeta^*$ and
   $\chi\chi^*\rightarrow\nu_R\nu_R$.}
 \label{fig: Som}
\end{figure}

When the DM particles froze out 
in the early universe, they became non-relativistic and the
non-perturbative Sommerfeld enhancement effect becomes important~\cite{
  Hisano:2004,Hisano:2005,Hisano:2006nn,AFSW,Chun1,Chun2,Chun3}.
We present the schematic diagrams of annihilation processes
with the Sommerfeld enhancement effect for
$\chi\chi^*\rightarrow \zeta\zeta^*$ in Figs.~\ref{fig: Som}(a,b)
and for $\chi\chi\rightarrow\nu_R\nu_R$ in Figs.~\ref{fig: Som}(c).
In fact, Sommerfeld effect contains an infinite series of the ladder diagrams.

\begin{figure}[htbp]
\centering
\captionsetup{justification=raggedright}
\subfigure[]{
 \includegraphics[width=0.48\textwidth,height=0.15\textheight]{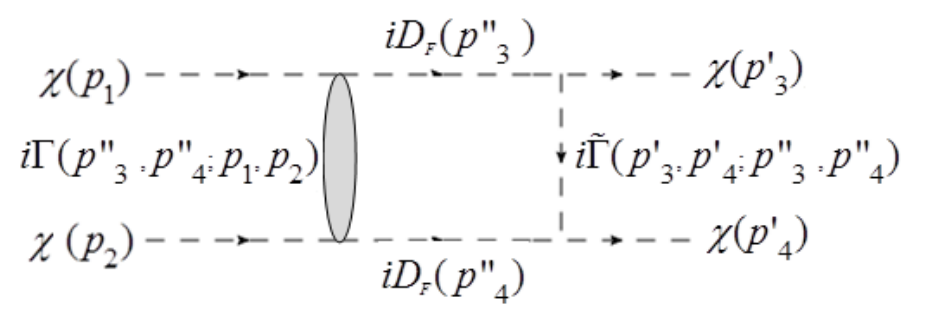}
}
\subfigure[]{
  \includegraphics[width=0.48\textwidth,height=0.15\textheight]{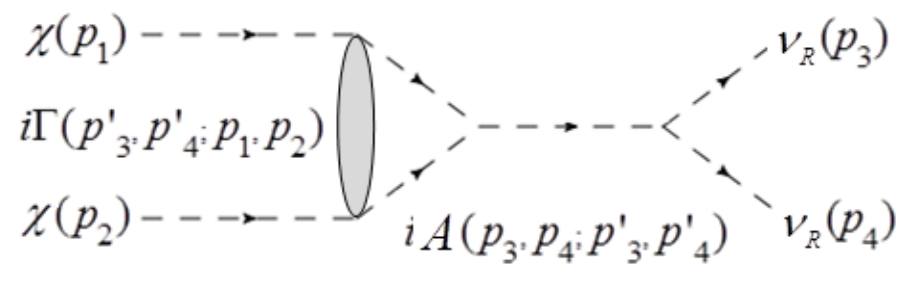}
}
\caption{Diagrams for (a) rescattering of $\chi\chi\to\chi\chi$, and
  (b) nonperturbative annihilation of $\chi\chi\to\nu_R\nu_R$.}
\label{fig: Som2}
\end{figure}

Let us first consider the easier case of Sommerfeld enhancement effect
on the annihilation process $\chi(p_1)\chi(p_2) \to \nu_R(p_3)\nu_R(p_4)$
through the exchange of the mediator $X = H$,
as shown in Fig.~\ref{fig: Som}(c).
When DM particles $\chi\chi$ become non-relativistic, they rescatter
off each other [Fig.~\ref{fig: Som2}(a)] before annihilating to $\nu_R\nu_R$.
The Sommerfeld enhanced amplitude $iA_{S}$  of
$\chi\chi\to \nu_R\nu_R$ annihilation process [Fig.~\ref{fig: Som2}(b)]
can be expressed as
\be
iA_{\mathcal{S}}(p_3, p_4, p_1, p_2)
&=&iA(p_3, p_4, p_1, p_2) \non\\
&+&\int \frac{d^4p'_3}{(2\pi)^4}
iA(p_3,p_4;p'_3,p'_4)
(iD_{F}(p'_3))
i\Gamma(p'_3,p'_4;p_1,p_2)
(iD_{F}(p'_4)) ,
\label{eq:wholeM1}
\en
where $iA$ is the amplitude of the annihilation process at tree level,
$D_{F}$ is the leptonic scalar DM propagator and $i\Gamma$ is the
amputated non-perturbative 4-point vertex function [Fig.~\ref{fig: Som2}(a)]
describing the rescattering process of
$\chi(p_1)\chi(p_2)\to \chi(p'_3)\chi(p'_4)$
and satisfying the following equation:
\begin{eqnarray}
i\Gamma(p'_3, p'_4, p_1, p_2) 
&=&
i\tilde\Gamma(p'_3, p'_4; p_1, p_2)\nonumber\\
&+&\int \frac{d^4p''_3}{(2\pi)^4}
i\tilde\Gamma(p'_3,p'_4;p''_3,p''_4)
\left[iD_{F}(p''_3)\right]
i\Gamma(p_1,p_2;p''_3,p''_4)
\left[iD_{F}(p''_4)\right].
\label{eq: master0}
\end{eqnarray}
In the above, we have the lowest order perturbative 4-point vertex
function given by
\be
i\tilde\Gamma(p'_3,p'_4;p''_3,p''_4)=-ig_X^2\frac{1}{(p''_3-p'_3)^2-m^2_X},
\en 
where $X$ is the mediator particle ($X=H$), and $p''_4=-p''_3+p'_3+p'_4$.
  Note that the dimensionless coupling strength of DM $\chi$ with the
  mediator $X$ is $g_X = g'_X/(2m_\chi)$, where the dimensionful coupling
  $g'_X$ is usually defined in the interaction Lagrangian.
  When the process is mediated by the Higgs boson ($H$), we have
$g_H = g'_H/(2m_\chi) = \lambda_{01}{\rm v}/(2m_\chi)$,
where the dimensionful coupling $g'_H = \lambda_{01}{\rm v}$ 
as described in the scalar potential [Eq. (1)].

Form Appendix~\ref{sec:Bethe_Salpeter}, we find that the pair of DM particles
$\chi\chi$ form a bound state and the  wave function $\psi(\vec r)$
satisfies the following Schr\"odinger equation:
\be
-\frac{1}{2\mu} \nabla^2\psi(\vec r)+V(\vec r)\psi(\vec r)
= E\psi(\vec r)=\frac{1}{2} \mu v^2\psi(\vec r) \, .
\label{eq: Schrodinger eq}
\en
where $\mu = m_\chi/2$ is the reduce mass of the bound state ($\chi\chi$),  and
$E=|\vec p|^2/2\mu\equiv \mu v^2/2$ is the total kinetic energy with
the relative velocity $v = v_{\rm lab}$ defined in
Eq.~(\ref{eq:vlab}).  $V(r)$ is a Yukawa-type potential
\be
V(r)=-\alpha_{X}\frac{e^{-m_Xr}}{r},
\label{eq: Yukawa}
\en
where $\alpha_{X}=g_X^2/4\pi$ and $m_X$ is the mass of mediator.

For the case of s-wave rescattering, the Sommerfeld
enhanced amplitude $iA_\mathcal{S}$ can be written as 
\be
iA_\mathcal{S}(p_1, p_2; p_3, p_4)=
iA(\vec p_1,\vec p_2;\vec p_3,\vec p_4) 
\psi(\vec r=0),
\label{eq: s wave scattering main text} 
\en
where $iA$ is the amplitude at tree level as explained in Appendix A.2.
Consequently, 
the Sommerfeld enhanced velocity averaged annihilation cross section
$\langle\sigma v\rangle_\mathcal{S} \simeq a_\mathcal{S}+b_\mathcal{S} v^2$ can be further
simplified as
\be
\langle\sigma v\rangle_\mathcal{S} \sim a_\mathcal{S} = \langle a \mathcal{S}(v)\rangle \, , \quad
\mathcal{\mathcal{S}} = |\psi_{l=0}(\vec r=0)|^2,
\en
where $\mathcal{S}$ is just the s-wave Sommerfeld enhancement factor (see Appendix A.2).

It is well known that there is no analytical solution with a Yukawa
potential in Eq.~(\ref{eq:  Schrodinger eq}), but the Hulth\'{e}n
potential maintains the same short and long distance behavior of the
Yukawa potential and has an analytical solution for s-wave function.
Hence it is a good approximation to employ the Hulth\'{e}n potential 
to obtain $|\psi_{l=0}(\vec{r}=0)|$ with a Yukawa potential~\cite{Cassel}, 
\be
V(\vec r) \simeq -\alpha_X
 \frac{(\pi^2 m_X/6) e^{-\pi^2 m_X r/6}}{1-e^{-\pi^2 m_X r/6}} \, ,
\label{eq: Huthen}
\en
where the mediator particle $X = H$. 
One can obtain the wave function~\cite{ChuaWong2017} in terms of Gamma
function $\Gamma$
\be
\psi_{l=0}(\vec r=0)&=&
i\frac{\pi^2\epsilon_X/6}{2\epsilon_v} \,
\Gamma\left(1-i \frac{\epsilon_v}{\pi^2\epsilon_X/6}\left(1+\sqrt{1-\frac{\pi^2\epsilon_X/6}{\epsilon^2_v}}\right)\right)
\non\\
&&\times
\Gamma\left(1-i \frac{\epsilon_v}{\pi^2\epsilon_X/6}\left(1-\sqrt{1-\frac{\pi^2\epsilon_X/6}{\epsilon^2_v}}\right)\right)
\bigg/\Gamma\left(\frac{-2i\epsilon_v}{\pi^2\epsilon_X/6}\right),
\label{eq: psi0Hulthen}
\en
with~\footnote{Note that the $\beta$ in the formula of~\cite{Cassel}
  is in fact $v/2$ in this work.}
\be 
\epsilon_v\equiv\frac{v}{2\alpha_{X}}, 
\quad
\epsilon_X\equiv\frac{m_X}{\alpha_{X} m_\chi},
\en 
and 
the s-wave Sommerfeld factor is given by~\cite{Feng:2010zp}
\be
\mathcal{S}(m_\chi, m_X, \alpha_X, v)=
|\psi_{l=0}(\vec r=0)|^2
=\frac{\pi}{\epsilon_v}
\frac{\sinh\left(\frac{2\pi\epsilon_v}{\pi^2\epsilon_X/6}\right)}
{\cosh\left(\frac{2\pi\epsilon_v}{\pi^2\epsilon_X/6}\right)
-\cos\left(2\pi \sqrt{\frac{1}{\pi^2\epsilon_X/6}-\frac{\epsilon^2_v}{(\pi^2\epsilon_X/6)^2}}\right)}.
\label{eq:Sold} 
\en 
For $\chi\chi\rightarrow\nu_R\nu_R$ annihilation process, the mediator particle is $X=H$. 
$\psi_{l=0}(\vec r=0)$ indeed goes to 1 in the $\alpha_X=0$ limit.
We will see that the analytic solution $\mathcal{S}=\psi_{l=0}(\vec r=0)$  agrees well with that obtained from numerically solving the Schr\"{o}dinger equation with the Yukawa potential.

When the mediator mass can be neglected,
the Yukawa potential can be approximated by a Coulomb potential:
\be
V(r)
\simeq-\frac{\alpha_{X}}{r}.
\label{eq: Coulomb}
\en
The corresponding s-wave function is given by~\cite{Coulomb}
\be
\psi_{l=0}^{(coul)}(\vec r)=\Gamma(1+i\gamma) e^{-\pi\gamma/2} e^{i\vec p\cdot\vec r}
{}_1\mathcal{F}_1(-i\gamma, 1,ipr-i\vec p\cdot \vec r),
\en
where ${}_1\mathcal{F}_1$ is the confluent hyper-geometric function of the first kind, and 
\be
\gamma=\frac{\alpha_{X} }{v}=\frac{\alpha_{X} \mu}{|\vec p|}.
\label{eq: gamma}
\en
In this approximation we have
\be
\psi_{l=0}^{(coul)}(\vec r=0)=\Gamma(1+i\gamma) e^{-\pi\gamma/2}.
\label{eq: psi coulomb}
\en 
Accordingly, the corresponding s-wave Sommerfeld factor in the Coulomb potential is
\be
\mathcal{S}^{(coul)}=|\psi_{l=0}^{(coul)}(\vec r=0)|^2
=\Gamma(1+i\gamma)\Gamma(1-i\gamma) e^{-\pi\gamma}
=\frac{2\pi\gamma}{e^{2\pi\gamma}-1}.
\label{eq: SQED}
\en
In fact $|\psi_{l=0}(\vec r=0)|$ goes to $|\psi_{l=0}^{Coul}(\vec r=0)|$
and the Sommerfeld factor $\mathcal{S}$ in Eq.~(\ref{eq:Sold}) does
reduce to $\mathcal{S}^{(coul)}$ in the large $m_\chi$ region~\cite{Cassel,Lebedev0}.



\begin{figure}[h!]
\centering
\captionsetup{justification=raggedright}
\subfigure[Analytical and numerical solutions.]{
  \includegraphics[width=0.45\textwidth]  {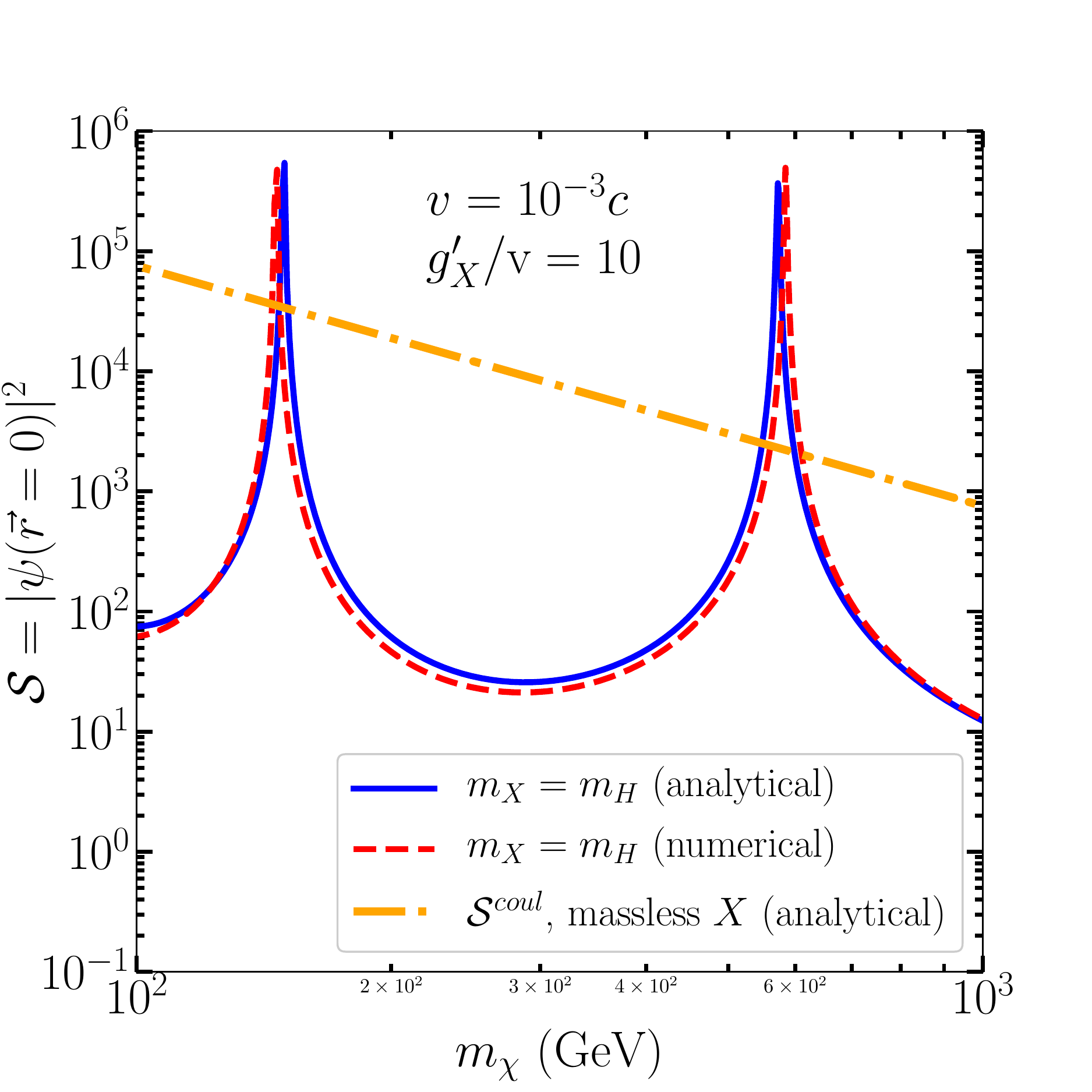}
}
\subfigure[Three different benchmark masses $m_X$.]{
 \includegraphics[width=0.45\textwidth]  {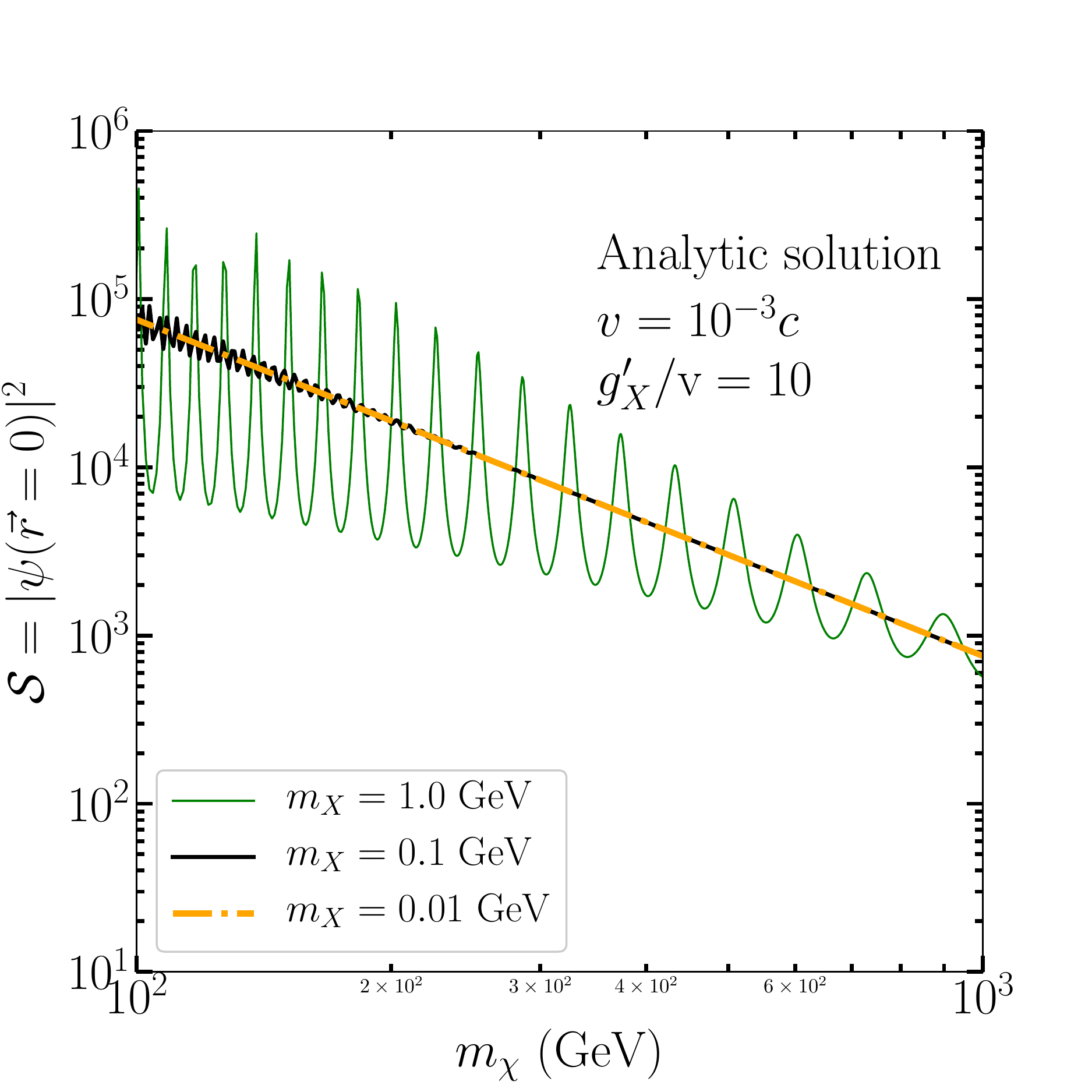}
}
\subfigure[Analytical and numerical solutions.]{
 \includegraphics[width=0.45\textwidth]  {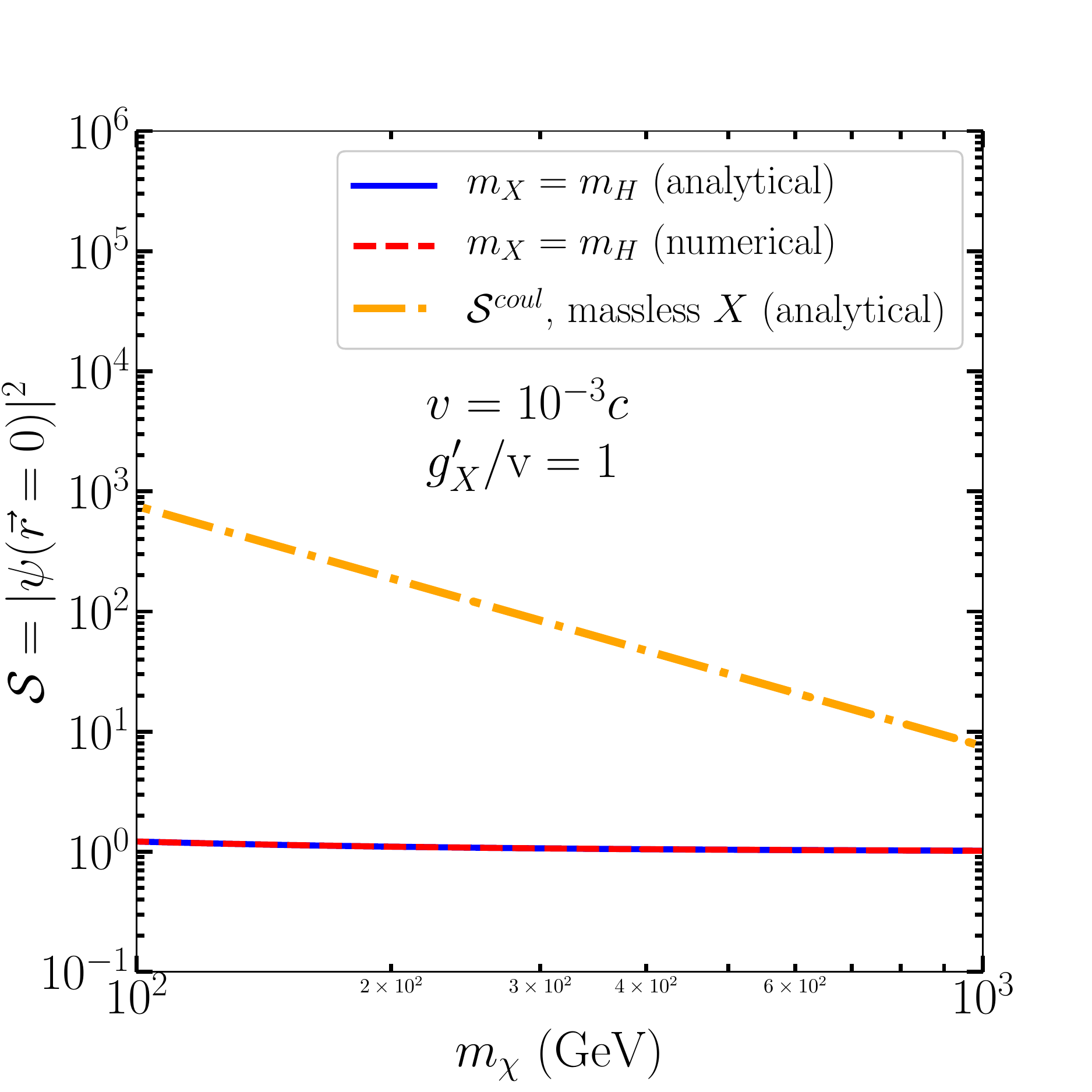}
}
\subfigure[Three different benchmark masses $m_X$.]{
 \includegraphics[width=0.45\textwidth]  {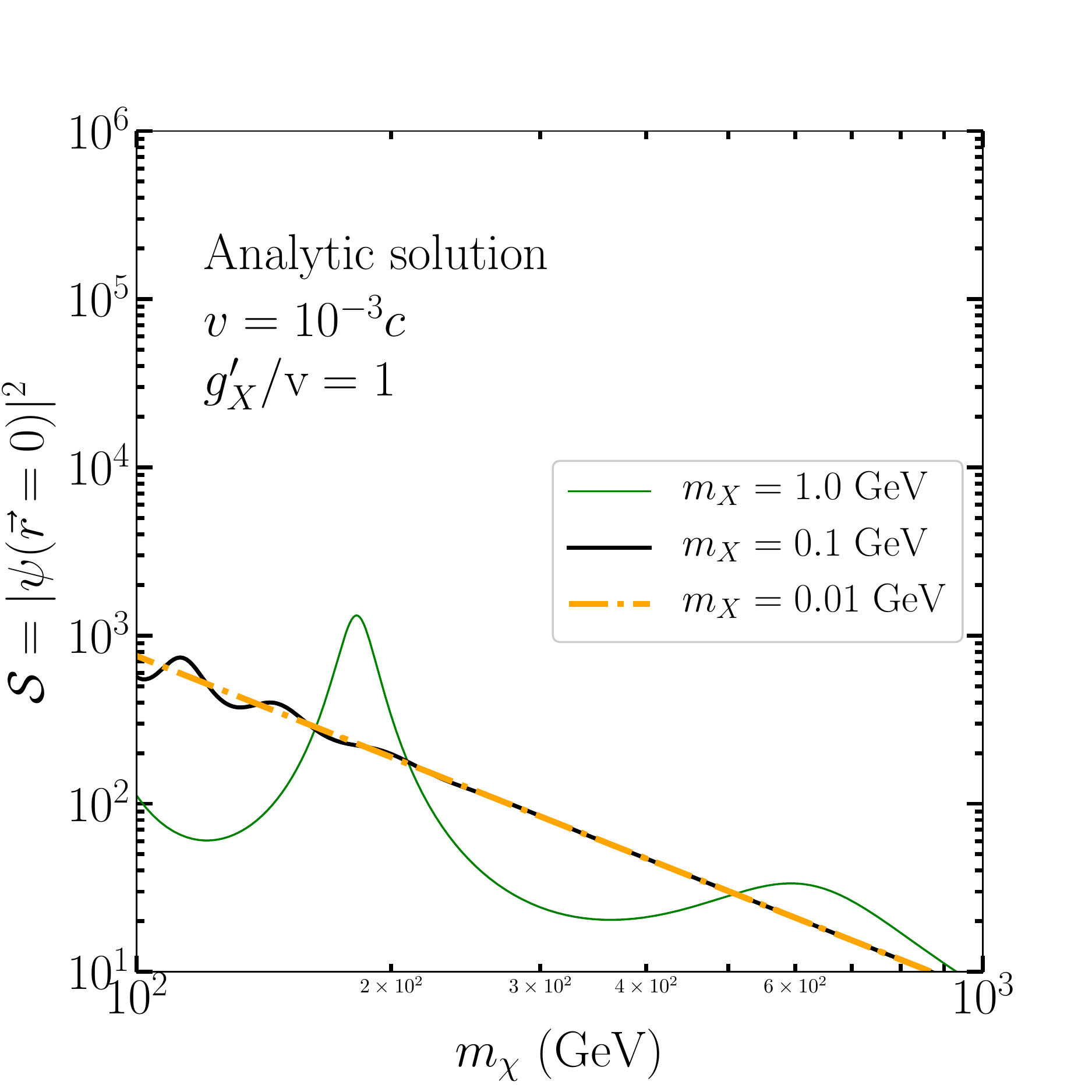}
}
\caption{
Sommerfeld factor $\mathcal{S}$ versus $m_\chi$ for the
present universe with $v=10^{-3}c$.  
The $g'_X$ parameter is the dimensionful coupling between the DM ($\chi$)
and the mediator $X$.
The left panels [(a) and (c)] present 
the analytical results (blue solid) with the Hulth\'en potential
[Eq. (\ref{eq: Huthen})] and $m_X = m_H$,
as well as numerical results (red dashed) with Yukawa potential
[Eq. (\ref{eq: Yukawa})] and $m_X = m_H$. 
For comparison, the analytical solution in Coulomb potential approximation 
is depicted by orange dashed lines. 
In the right panels [(b) and (d)], the analytical
solutions in Huth\'en potential approximation are shown with
three values of the mediator mass $m_X = 0.01$ (orange dot-dashed),
0.1 (black solid), and 1 $\gev$ (green solid).
\label{fig: Som gX}
 }
\end{figure}

As mentioned above, we only keep the first term (the $a$ term) in
Eqs.(\ref{eq: sigma v1}-\ref{eq: sigma v5}) for both relic density
calculation and the indirect annihilation processes, namely, the
s-wave contribution,
with $\bra \sigma_{\mathrm{ann}}v \ket \approx a + {\cal O}(v^2)$.
Hence we show the $a$-term for each annihilation process as follows:
\be
a^{\zeta\zeta^*}
&=&\frac{\sqrt{m_\chi^2-m_\zeta^2}}{32 \pi m_\chi^3}
\left\{ \frac
{(\mu_{12}^2-\lambda_{12}m_\zeta^2+2\lambda_{12} m_\chi^2)^2}
{(2 m_\chi^2-m_\zeta^2)^2}
+\frac
{\lambda_{01}^2\lambda_{02}^2 v^4}
{[(4 m_\chi^2-m_H^2)^2+m_H^2 \Gamma_H^{2}]}\right.\non\\
&-&\left.\frac
{2\lambda_{01}\lambda_{02} v^2(4m_\chi^2-m_H^2)(4\mu_{12}^2-\lambda_{12} m_\zeta^2+2 \lambda_{12} m_\chi^2)}{(2 m_\chi^2-m_\zeta^2)[(4 m_\chi^2-m_H^2)^2+m_H^2 \Gamma_H^2]}\right\},\\
a^{f\bar f}&=&\frac{C_f\lambda_{01}^2 m_f^2(m_\chi^2-m_f^2)^{3/2}}{4\pi m_\chi^3[(4 m_\chi^2-m_H^2)^2+m_H^2 \Gamma_H^2]},
\label{eq: a1}
\en
\be
a^{VV}&=&\frac{\lambda_{01}^2\sqrt{m_\chi^2-m_W^2}(4m\chi^4-4m_\chi^2m_W^2+3m_W^4)}{8\pi m_\chi^3((4 m_\chi^2-m_H^2)^2+m_H^2 \Gamma_H^2))}{S},
\\ \non\\
a^{\nu_{Ri}\nu_{Rj}}&=&\frac{\mu_{12}^2|f_{ij}|^2(2m_\chi^2-m_\nu^2)\sqrt{m_\chi^2-m_\nu^2}}{\pi m_\chi^3[(4m_\chi^2-m_\zeta^2)^2+m_\zeta^2 \Gamma_\zeta^2]},\\
a^{\zeta H}&=&\frac{\mu_{12}^2v^2}{128\pi m_\chi^4 D}
\left\{4\sqrt{A}\left[\lambda_{02}^2
+\frac{8 \lambda_{01}^2 D(A+4m_H^2m_\zeta^2)}{(A+4 m_H^2m_\zeta^2)^2}\right]\right.\non\\
&-&
\left. \frac{128\lambda_{01}m_\chi^2\left[\lambda_{02}(4m_\chi^2-m_\zeta^2)C-8\lambda_{01}m_\chi^2D\right]}{\sqrt{B}C}\right\}.
\label{eq: a5}
\en
For $\chi\chi\to \nu_R\nu_R$ and $\zeta H$ annihilation processes, we have  
\be
\langle\sigma v\rangle_\mathcal{S}(\chi\chi\to\nu_R\nu_R)
&\simeq&\langle  a^{\nu_R\nu_R} \mathcal{S}(m_\chi, m_H, \alpha_H, v)\rangle,
\\
\langle\sigma v\rangle_\mathcal{S}(\chi\chi\to\zeta H)
&\simeq&\langle a^{\zeta H} \mathcal{S}(m_\chi, m_H, \alpha_H, v)\rangle.
\en
Accordingly, we can analytically obtain the Sommerfeld factor 
$\mathcal{S}=|\psi_{l=0}(\vec{r}=0)|$ for the $\chi\chi\to\nu_R\nu_R$ and
$\chi\chi\to\zeta H$ annihilation processes.

Fig.~\ref{fig: Som gX} shows the s-wave Sommerfeld factor in the
  present universe with ($v= 10^{-3} c$) as a function of
  the DM mass ($m_\chi$) for two values of the coupling
  $g'_X/ {\rm v} = 10$ [(a) and (b)] as well as $g'_X/{\rm v} = 1$ [(c) and (d)].
The left panels [(a) and (c)] present 
the analytical results with the Hulth\'en potential
[Eq. (\ref{eq: Huthen})] and $m_X = m_H$,
as well as numerical results with Yukawa potential
[Eq. (\ref{eq: Yukawa})] and $m_X = m_H$. 
In addition, we show the analytical solution in Coulomb potential
approximation.
In the right panels [(b) and (d)], the analytical
solutions in Huth\'en potential approximation are presented
with three values of the mediator mass
$m_X = 0.01, 0.1$, and 1 $\gev$.
The Coulomb potential approximation in the left panels and the
anlytical solution with $m_X = 0.01\gev$ in the right panels
are both represented with orange dot-dashed lines  
because there is no difference between them numerically.

We find that all curves in Figs. 6(b) and 6(d)
   with a massive $m_X$ oscillate with an amplitude that
increases with an increasing mediator mass, while the frequency of
oscillation increases with a decreasing mediator mass. 
Comparing Fig.~\ref{fig: Som gX}(a,b) and Fig.~\ref{fig: Som gX}(c,d),
we see that the Sommerfeld enhancement factor $\mathcal{S}$ decreases
with a decreasing coupling $g'_X$.
Furthermore, the strength of Sommerfeld enhancement decreases with
an increasing $m_\chi$. The reason can be seen as follows:
\begin{itemize}
\item First, these curves 
  oscillate around the curve generated by the Coulomb potential.
\item The Coulomb potential approximation provides the central value
  for $\mathcal{S}$.
\item From Eq.~(\ref{eq: gamma}) and Eq.~(\ref{eq: SQED}),
 we note that $\mathcal{S}^{(coul)}$ is a function of
$\alpha_X=g^2_X/{4\pi}$. Hence the enhancement factor $\mathcal{S}$ 
 is suppressed by an increasing $m_\chi$ owing to $g_X=g'_X/m_\chi$.
\end{itemize}


\begin{figure}[t!]
\centering
\captionsetup{justification=raggedright}
{
\includegraphics[width=0.6\textwidth]{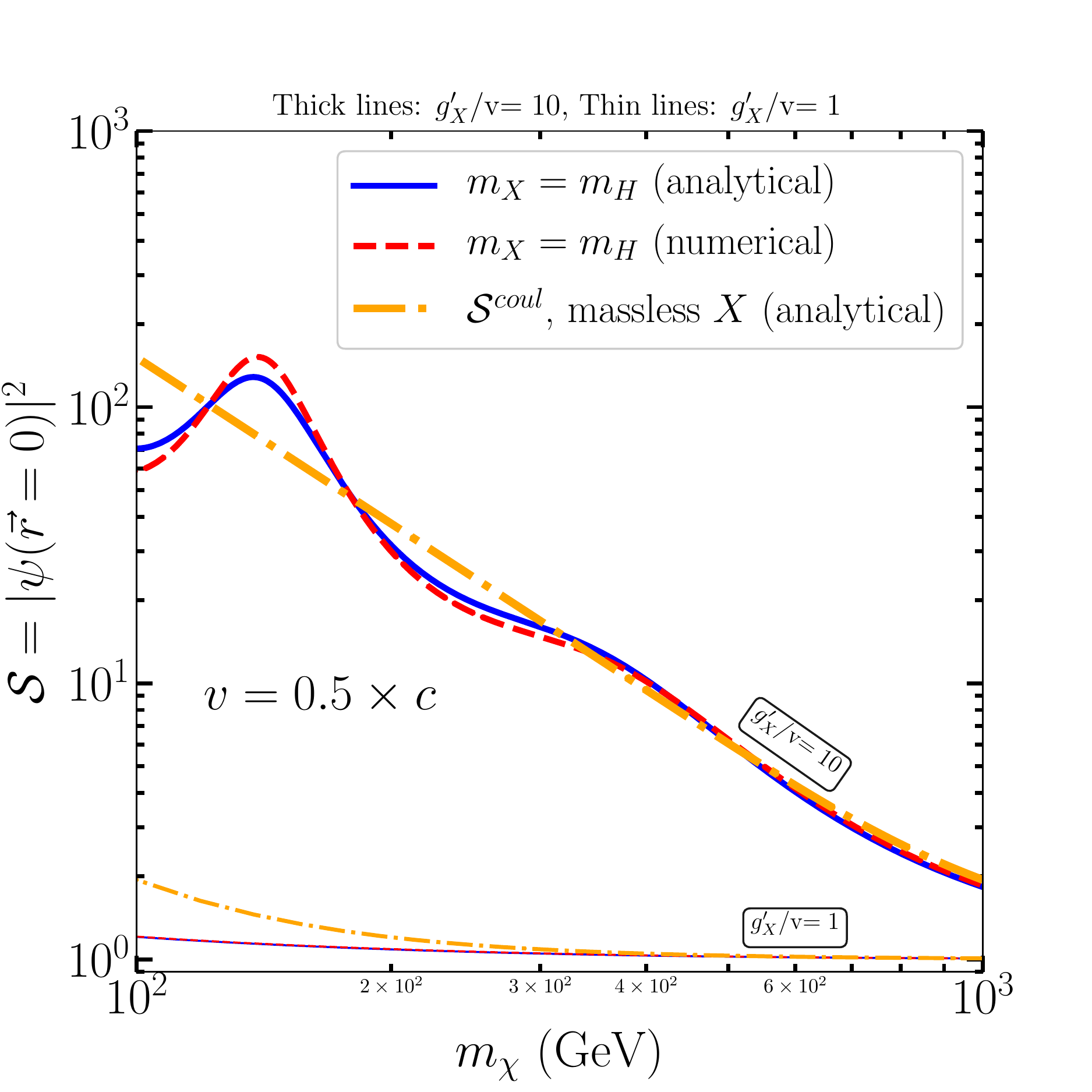}
}
\caption{Sommerfeld factor $\mathcal{S}$ versus $m_\chi$ for the early
  universe with $v=0.5\times c$.
The color scheme is same as Fig.~\ref{fig: Som gX}. 
The coupling $g^\prime_X/{\rm v}$ for the upper thick lines are fixed to $10$ 
while the value of the bottom thin lines are $1$.
\label{Fig:SE_halfc}}
\end{figure}

In Fig.~\ref{Fig:SE_halfc}, we show $\mathcal{S}$ at the early
universe with $(v=0.5\times c)$ as a function of $m_\chi$.
The colored scheme is the same as Fig.~\ref{fig: Som gX} 
while the thick and thin lines are based on $g^\prime_X/{\rm v}=10$
and $g^\prime_X/{\rm v}=1$, respectively.
Clearly, comparing with low-velocity DM in present universe as shown
in Fig.~\ref{fig: Som gX}, the nonperturbative effect for
high-velocity becomes much weaker resulting in much smaller Sommerfeld
enhancement in the early universe.



From Fig.~\ref{fig: Som gX} and 
Fig.~\ref{Fig:SE_halfc}, we find that, 
the larger coupling strength or smaller DM relative
velocity $v$ gives a greater value of the Sommerfeld factor.
Roughly speaking, the Sommerfeld factor used at the present stage
($v\simeq 10^{-3}c$) is 2$\sim$3 order of magnitude greater than that
in freeze-out stage $(v\simeq 0.5\times c)$ with a light mediator mass.
This main characteristics of velocity dependent SICS is used to solve
the small scale problem.

We have shown in Fig.~\ref{fig: Som gX} and Fig.~\ref{Fig:SE_halfc} that the numerical result
agrees well with the analytic solution in Hulth\'en approximation at
$m_X=m_H$.
Nevertheless, we see that it is impossible to obtain an
analytical form for $\mathcal{S}$ in the process of $\chi\chi^*\to\chi^*\chi$,
which simultaneously involves the exchanges of 
$H$ and $\zeta$ particles schematically shown in Fig.~\ref{fig: Som}(a,b). 
The potential then becomes the sum of two Yukawa-type potential
\be
V(r)=-\alpha_H\frac{e^{-m_Hr}}{r}-\alpha_\zeta\frac{e^{-m_\zeta r}}{r}.
\en
where
$\alpha_X={g_X^2}/{4\pi}, (X=H,\zeta)$, $g_H={\lambda_{01}{\rm v}}/{2m_\chi}$ 
and $g_\zeta={\mu_{12}}/{m_\chi}$. 
There is no analytical form for the Sommerfeld enhancement factor $\mathcal{S}$. 
Therefore, we develop a numerical solution for $\mathcal{S}$ as given in 
Appendix~\ref{sec:numS}.

\subsection{Numerical Results for Relic Density and Indirect Search}

In this subsection, we present our numerical results for the indirect
search and the relic density. For the indirect search,
we compare our theoretical results with the most stringent limits
from the Fermi-LAT \cite{Fermi-LAT2015,Fermi-LAT:2016uux} and the
H.E.S.S. results~\cite{HESS2016}.
Both Fermi-LAT and H.E.S.S. astrophysical observations do not show the
significant $\gamma$-ray signal above background.
Instead, Fermi-LAT provides upper limits on $\la \sigma_{\rm ann} v\ra$
for DM annihilating into $W^+W^-$ and the SM fermion pairs:
$b{\bar b}, u{\bar u}, \tau^+\tau^-, \mu^+\mu^-, e^+e^-$
at $95\%$ confidence level with WIMPs masses between 2 GeV to 10 TeV,
while H.E.S.S. gives the  upper limits on $\la \sigma_{\rm ann} v\ra$
for DM annihilating into $W^+W^-$ and the SM fermion pairs:
$ t{\bar t}, b{\bar b}, \tau^+\tau^-, \mu^+\mu^-$
with masses from 160 GeV to 70 TeV.

\begin{figure}[t!]
\centering
\captionsetup{justification=raggedright}
{
  \includegraphics[width=0.4\textwidth,height=0.15\textheight]  {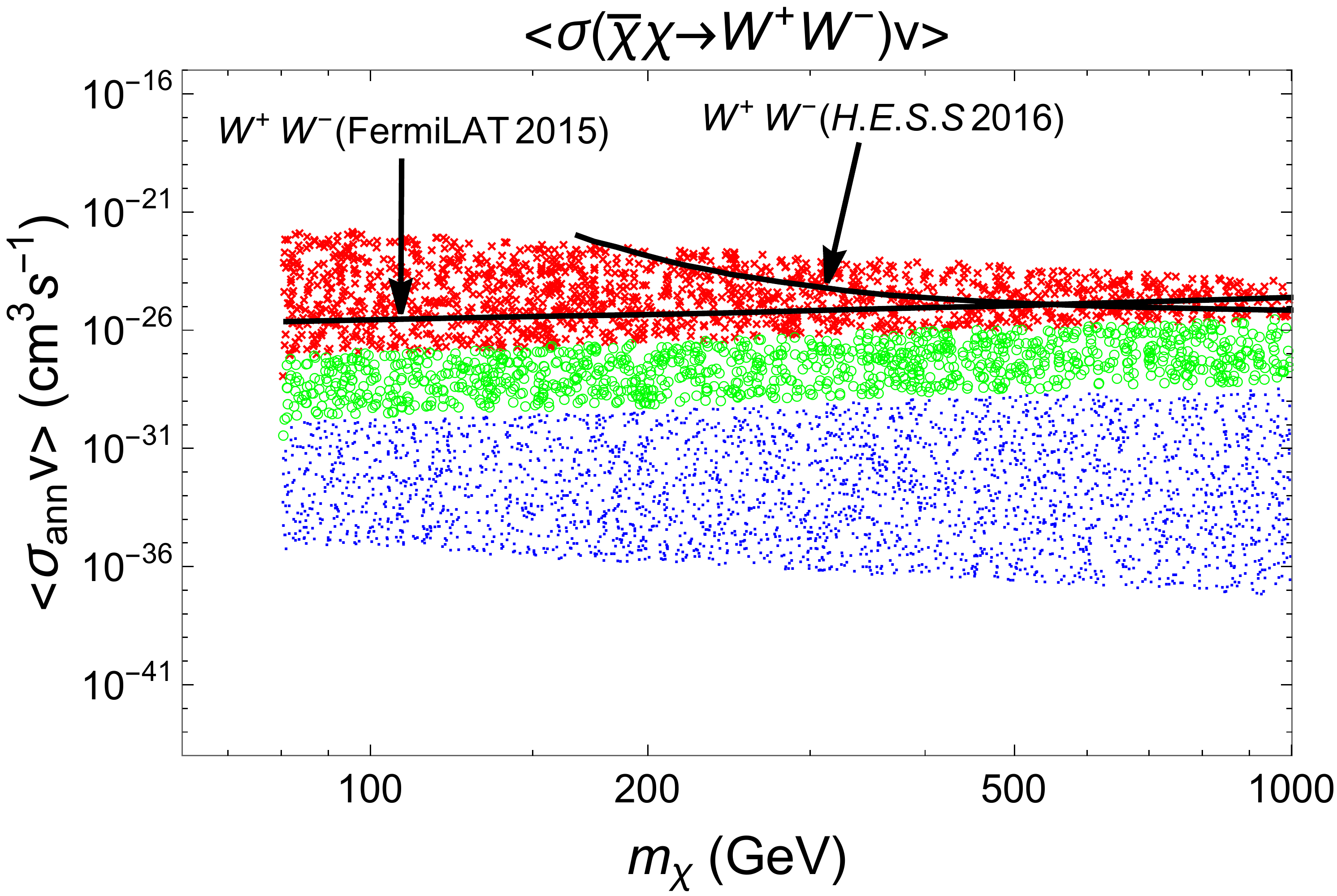}
}
{
 \includegraphics[width=0.4\textwidth,height=0.15\textheight]  {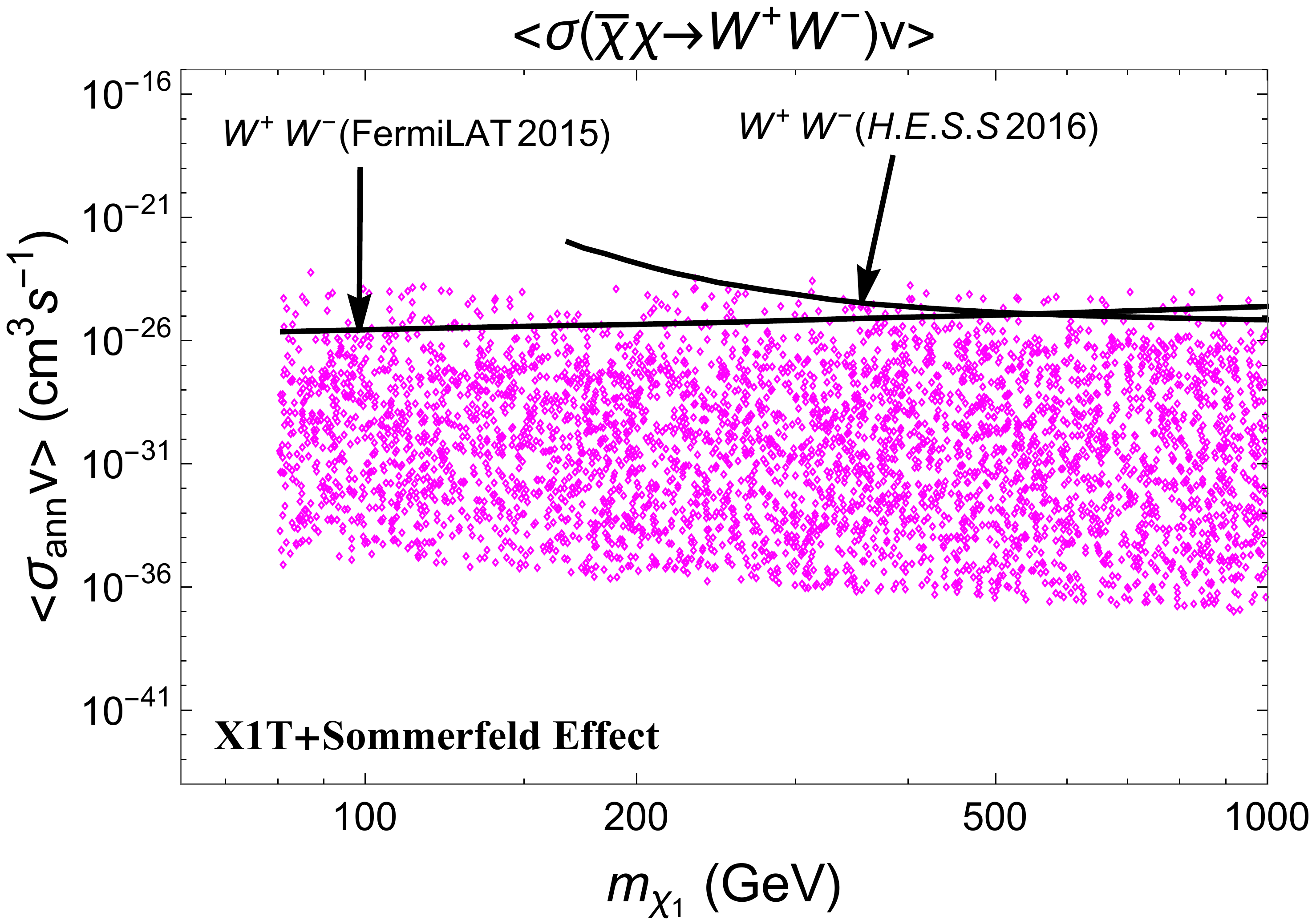}
}
{
 \includegraphics[width=0.4\textwidth,height=0.15\textheight]  {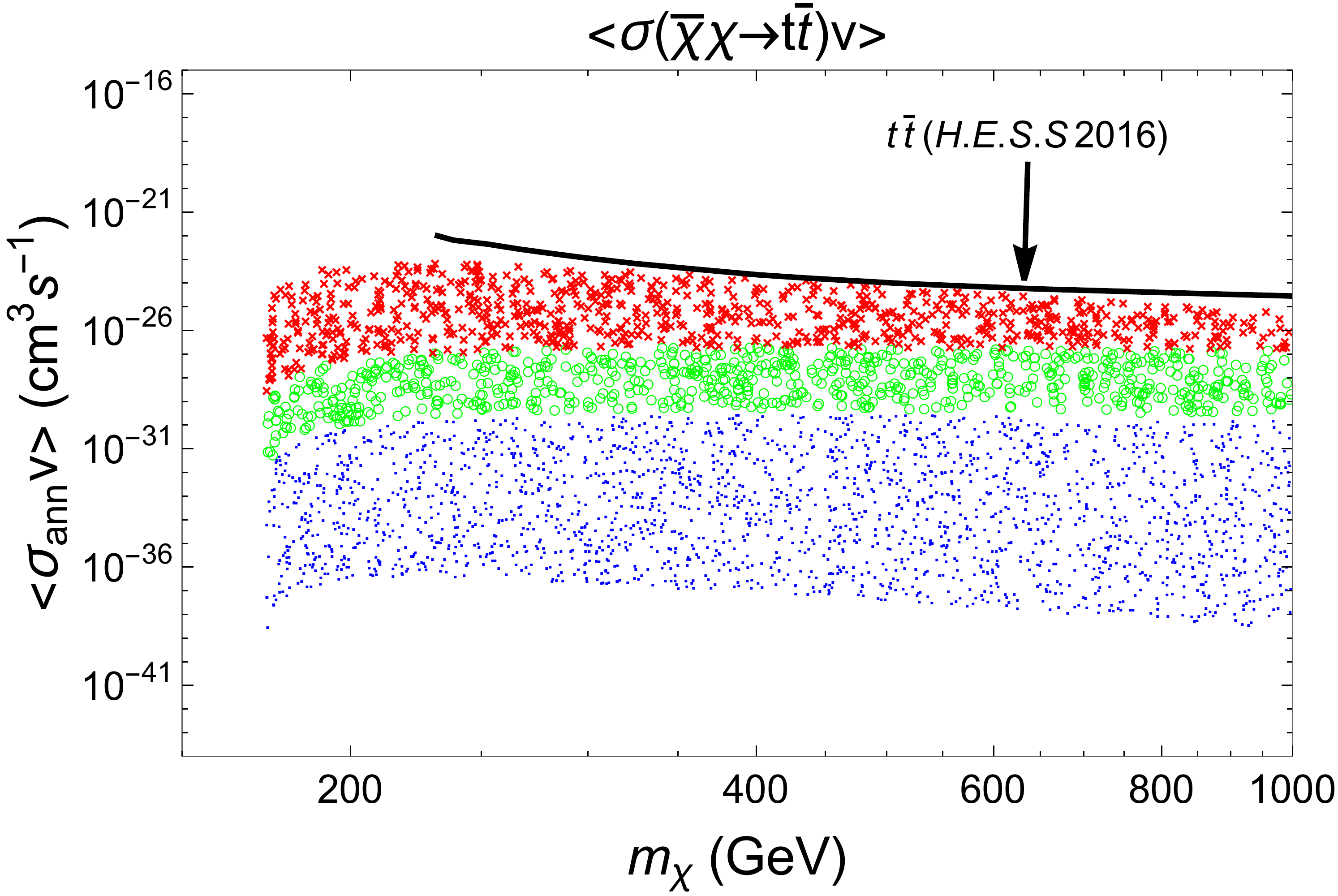}
}
{
 \includegraphics[width=0.4\textwidth,height=0.15\textheight]  {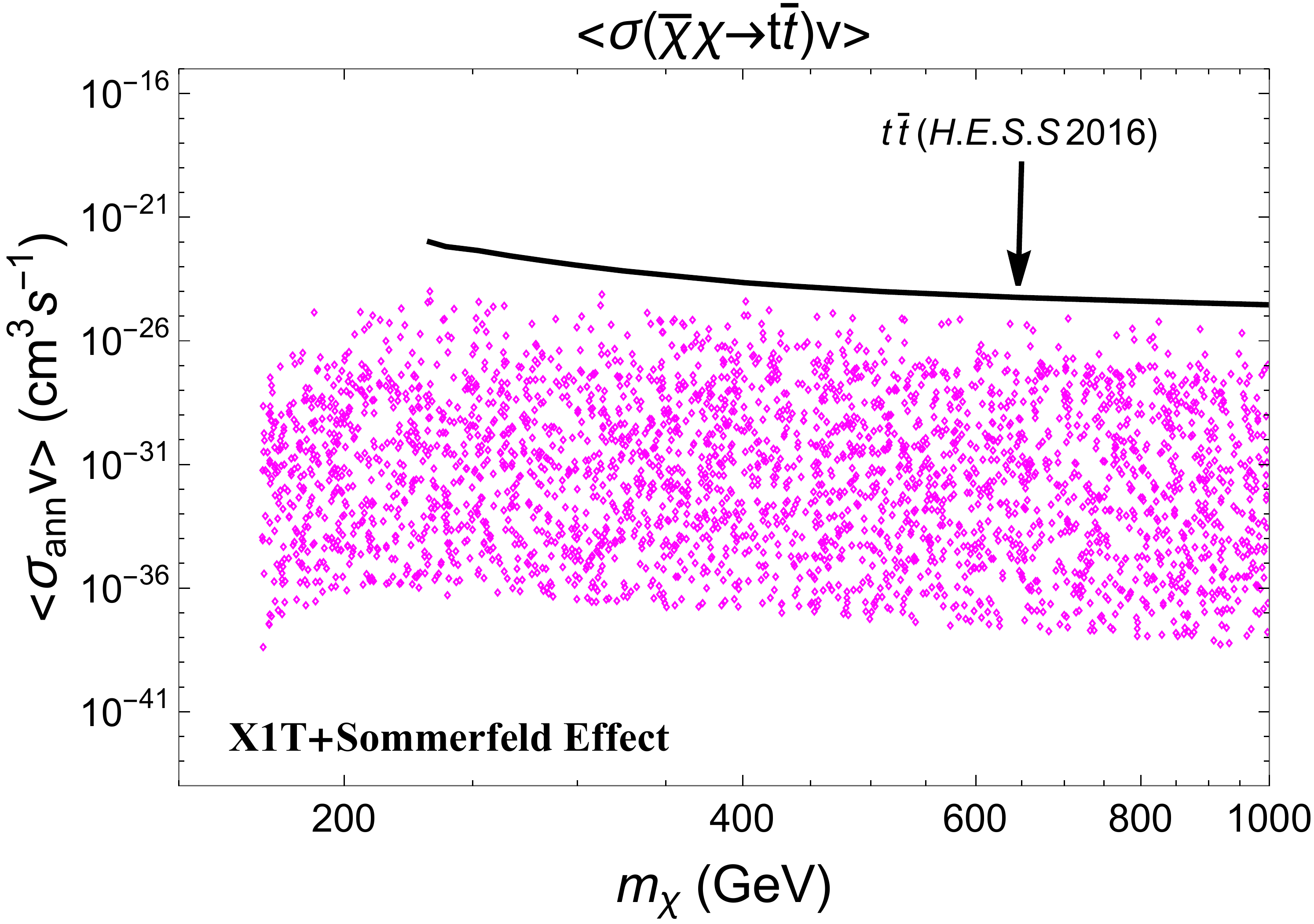}
}
{
 \includegraphics[width=0.4\textwidth,height=0.15\textheight]  {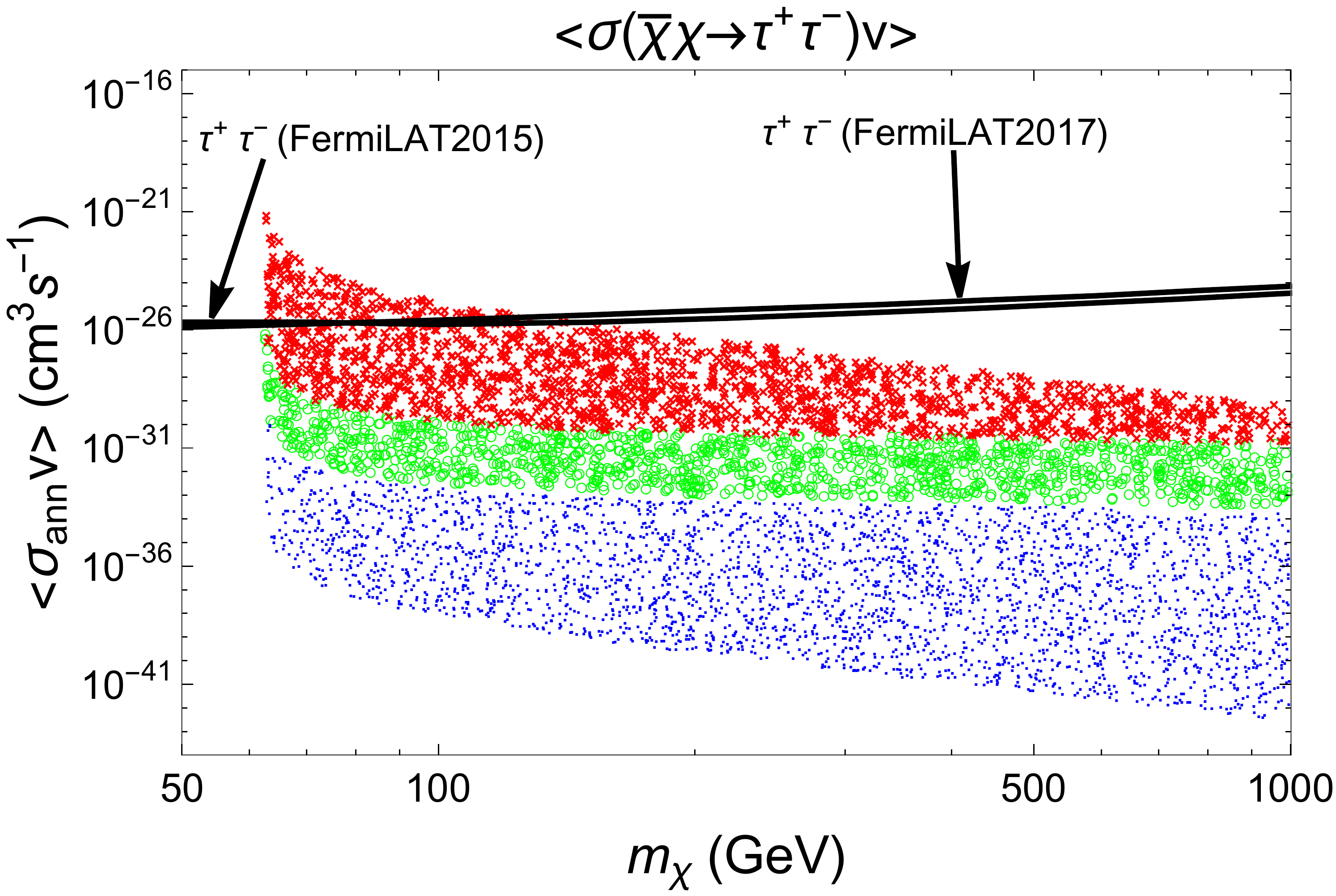}
}
{
 \includegraphics[width=0.4\textwidth,height=0.15\textheight]  {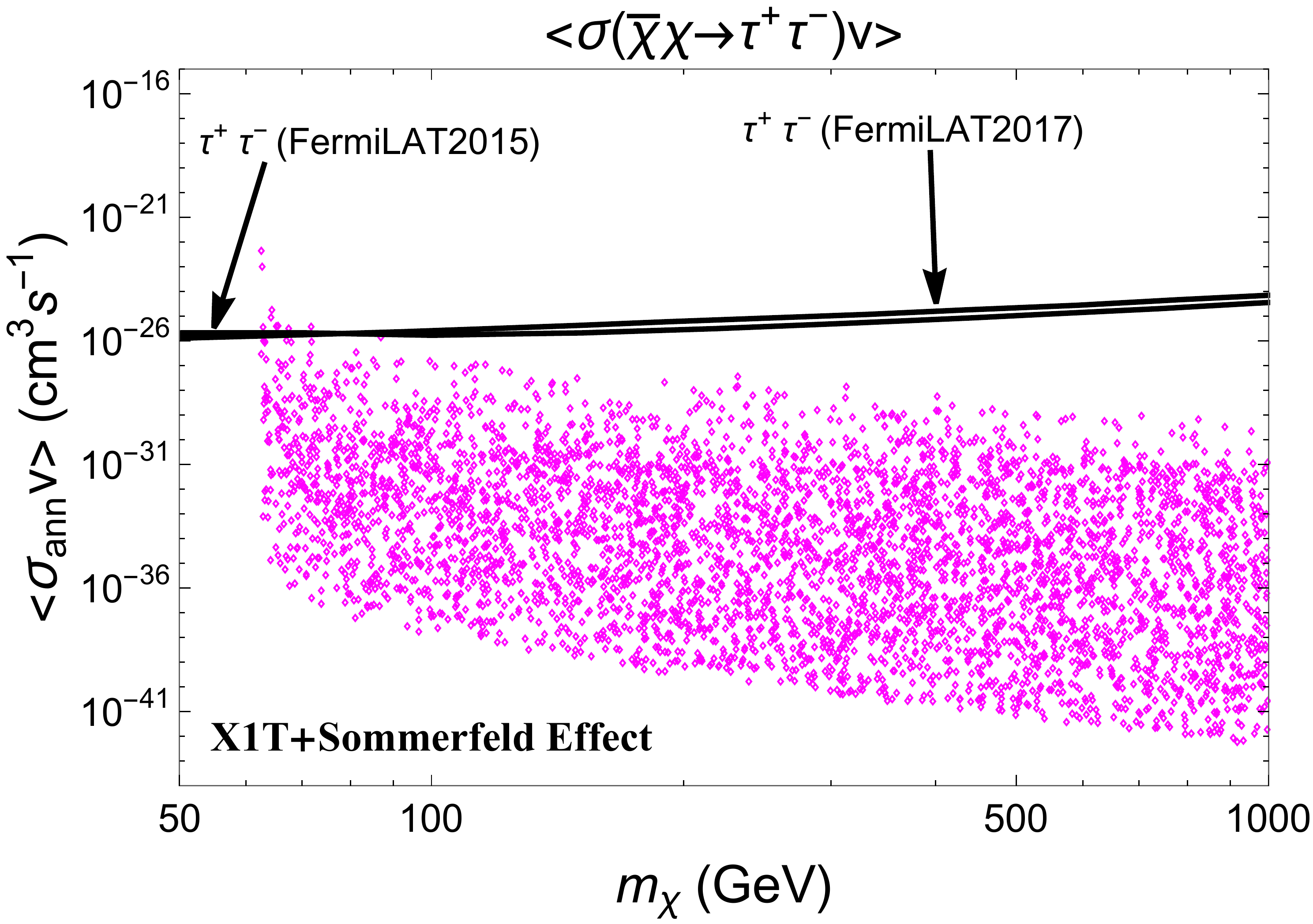}
}
{
 \includegraphics[width=0.4\textwidth,height=0.15\textheight]  {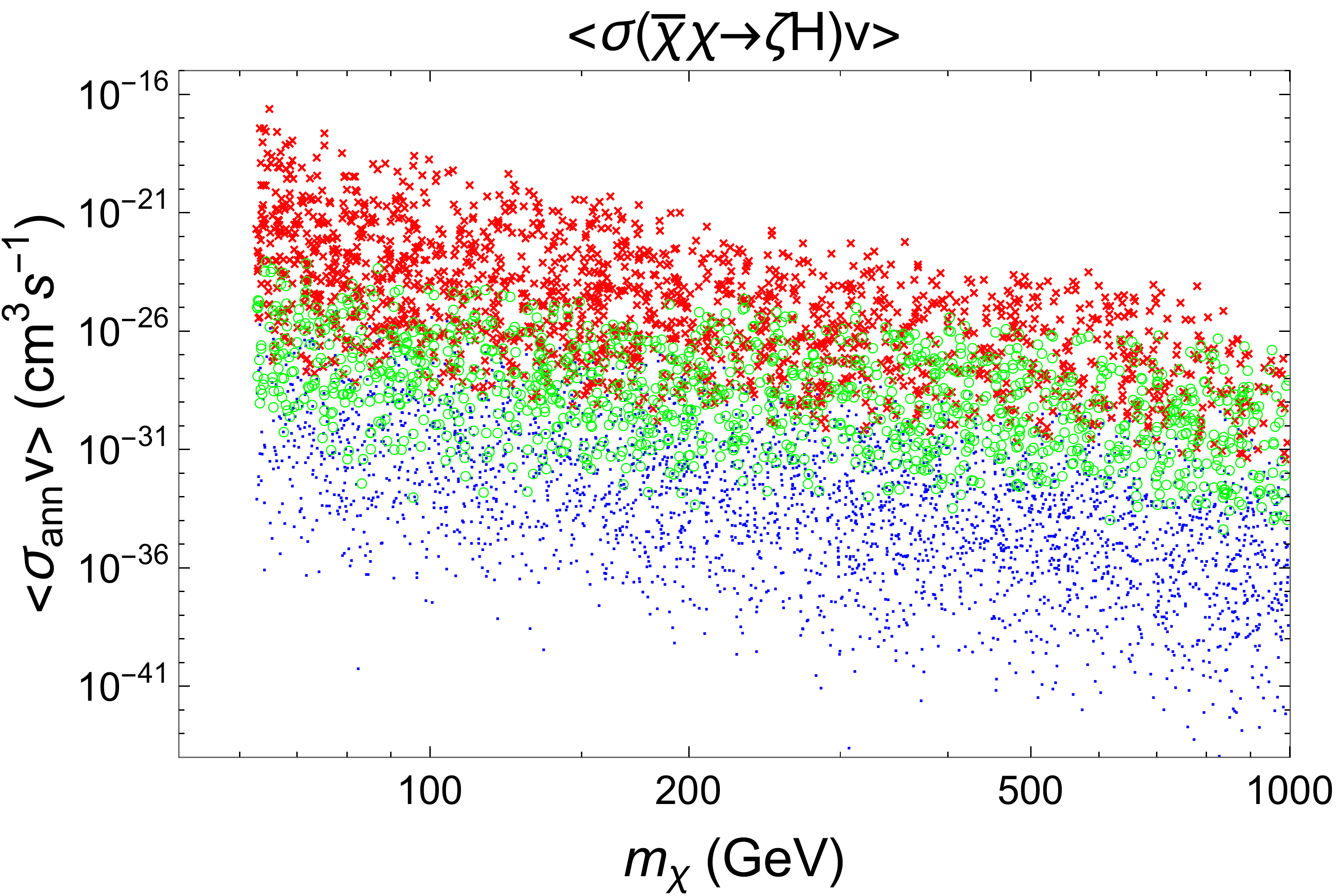}
}
{
 \includegraphics[width=0.4\textwidth,height=0.15\textheight]  {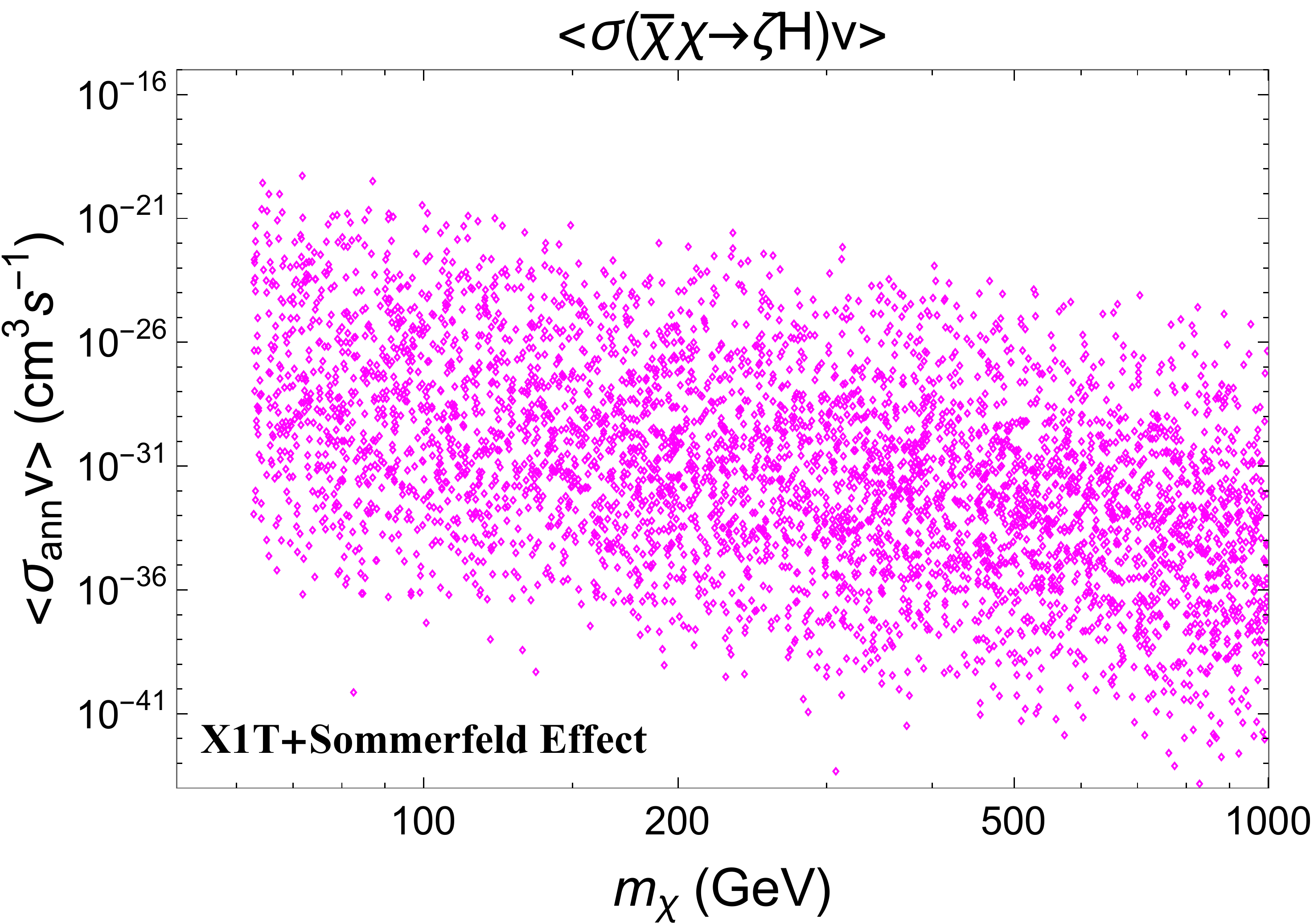}
}
{
 \includegraphics[width=0.4\textwidth,height=0.15\textheight]  {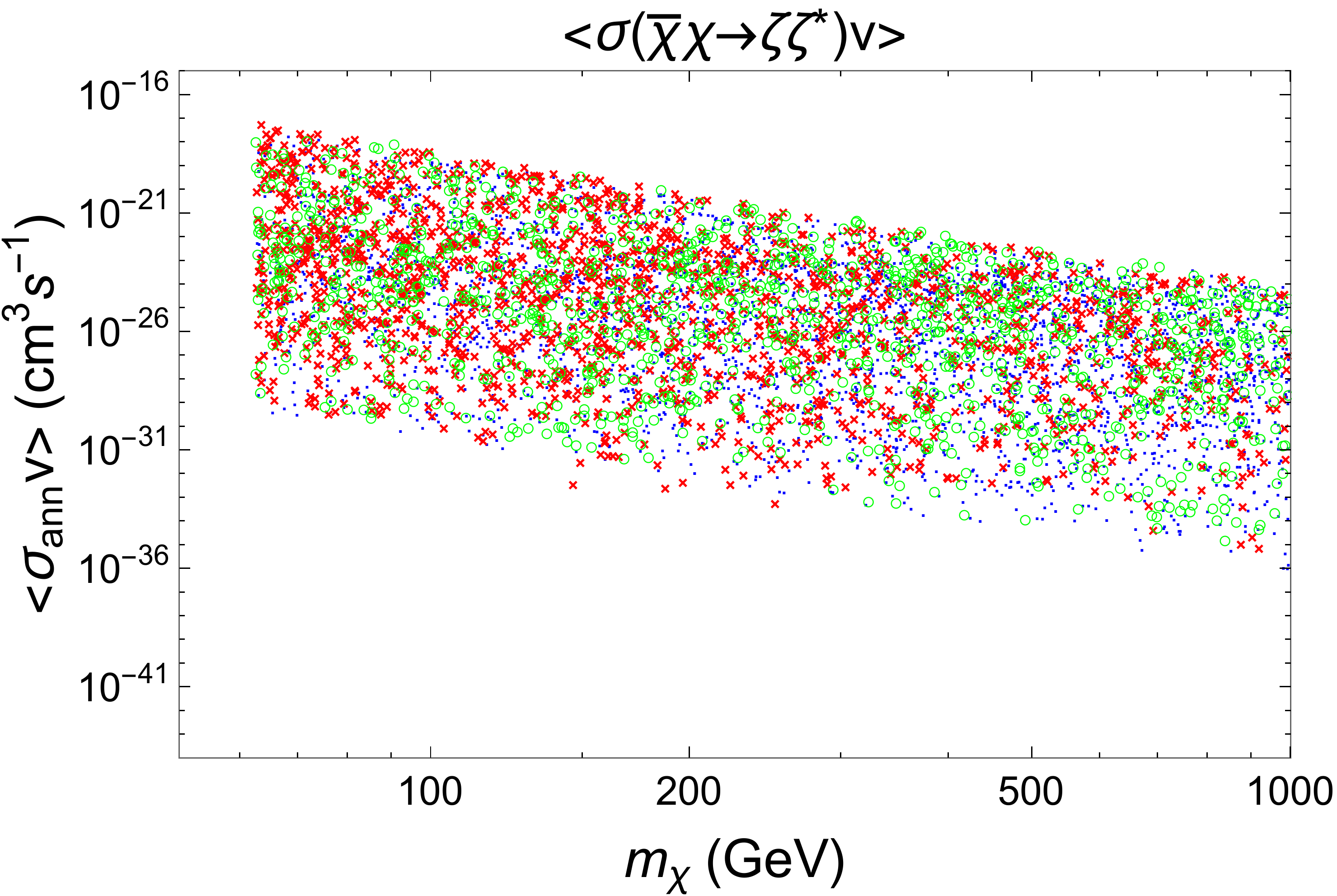}
}
{
 \includegraphics[width=0.4\textwidth,height=0.15\textheight]  {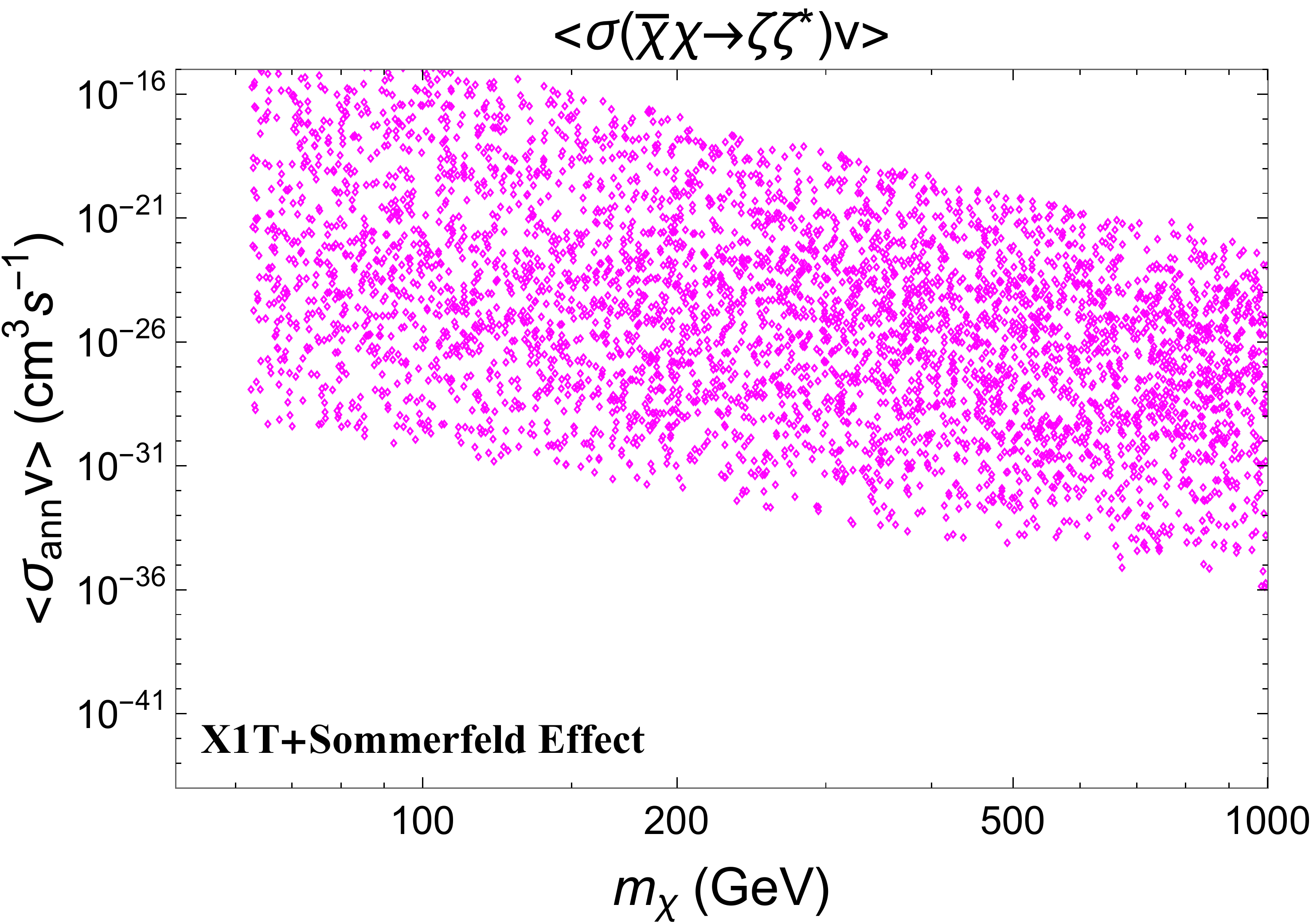}
}
\caption{Thermally averaged annihilation cross section times velocity for
    the leptonic scalar dark matter ($\chi$)
    $\langle\sigma_{\rm ann} v\rangle$ in different
    channels. [Left] Without the Sommerfeld effect:
  red ``${\color{red}\times}$", green ``${\color{green}\circ}$" and
  blue ``${\color{blue}\cdot}$" denote the scenario excluded by XENON1T, 
  testable soon by future underground detectors, and 
  below the neutrino floor, respectively.
  [Right] the Sommerfeld effect for those samples surviving from
  XENON1T limit, namely
  green ``${\color{green}\circ}$" and blue ``${\color{blue}\cdot}$"
  in the left panels.
    }
\label{fig: Indirect}
\end{figure}

Fig.~\ref{fig: Indirect} presents $\langle\sigma_{\rm ann} v\rangle$
for leptonic scalar DM ($\chi\chi$) annihilating into
$W^+W^-$, $t\bar t$, $\tau^+\tau^-$, $\zeta H$ and $\zeta \zeta^*$.
The samples above the upper limits of Fermi-LAT and H.E.S.S are ruled out.
The plots on the left-handed do not include the Sommerfeld effect but
the Sommerfeld effect are considered in the plots on the right-handed side. 
In each plot on the left-handed side, 
the same color scheme as presented in Fig.~\ref{fig: direct} is used. 
In the panels of the right column, 
we show the Sommerfeld effect for the data survived from XENON1T limits, 
namely the samples taken from those green ``${\color{green}\circ}$" and
blue ``${\color{blue}\cdot}$" in the left panels.

Without considering the Sommerfeld effect, we see that DM can only be
detected with $m_\chi\gtrsim 1$ TeV via $W^+W^-$ or
the Higgs resonance annihilation via $b\bar b$ and
$\tau^+\tau^-$ channel. 
Clearly, the cross sections can be enhanced by the Sommerfeld effect. 
  Albeit the enhancement differs from channel to channel, 
it is interesting that the cross section of the channel $\zeta\zeta^*$
is overall enhanced.
We see that DM annihilating to a pair of $\zeta\zeta^*$ is dominant 
while this channel is not detectable because $\zeta$ eventually decays
to $\nu_R$.

\begin{figure}[t!]
\centering
\captionsetup{justification=raggedright}
\subfigure[\ Sommerfeld effect not included]{
 \includegraphics[width=0.44\textwidth]  {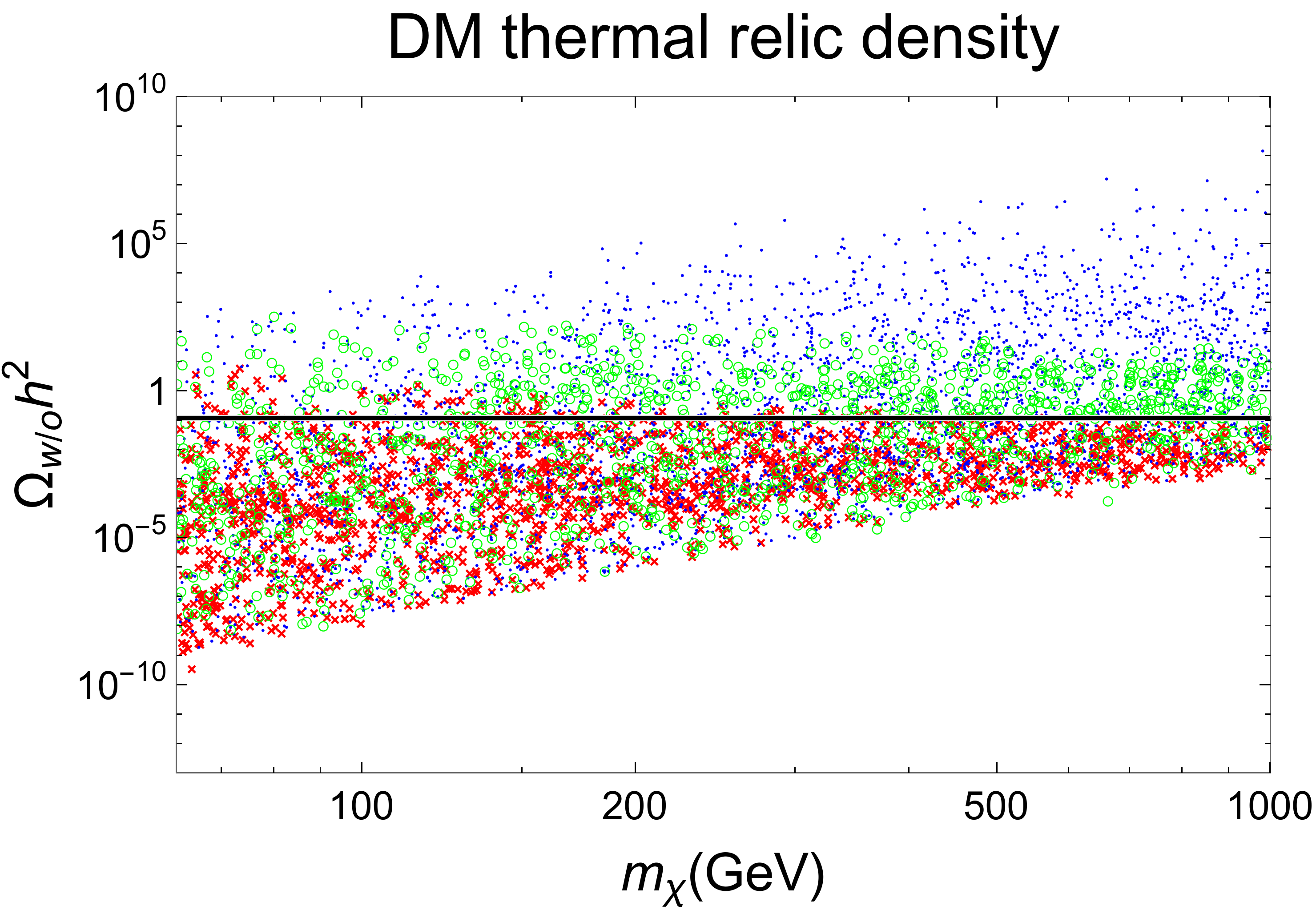} }
\subfigure[\ Sommerfeld effect included]{
 \includegraphics[width=0.44\textwidth]  {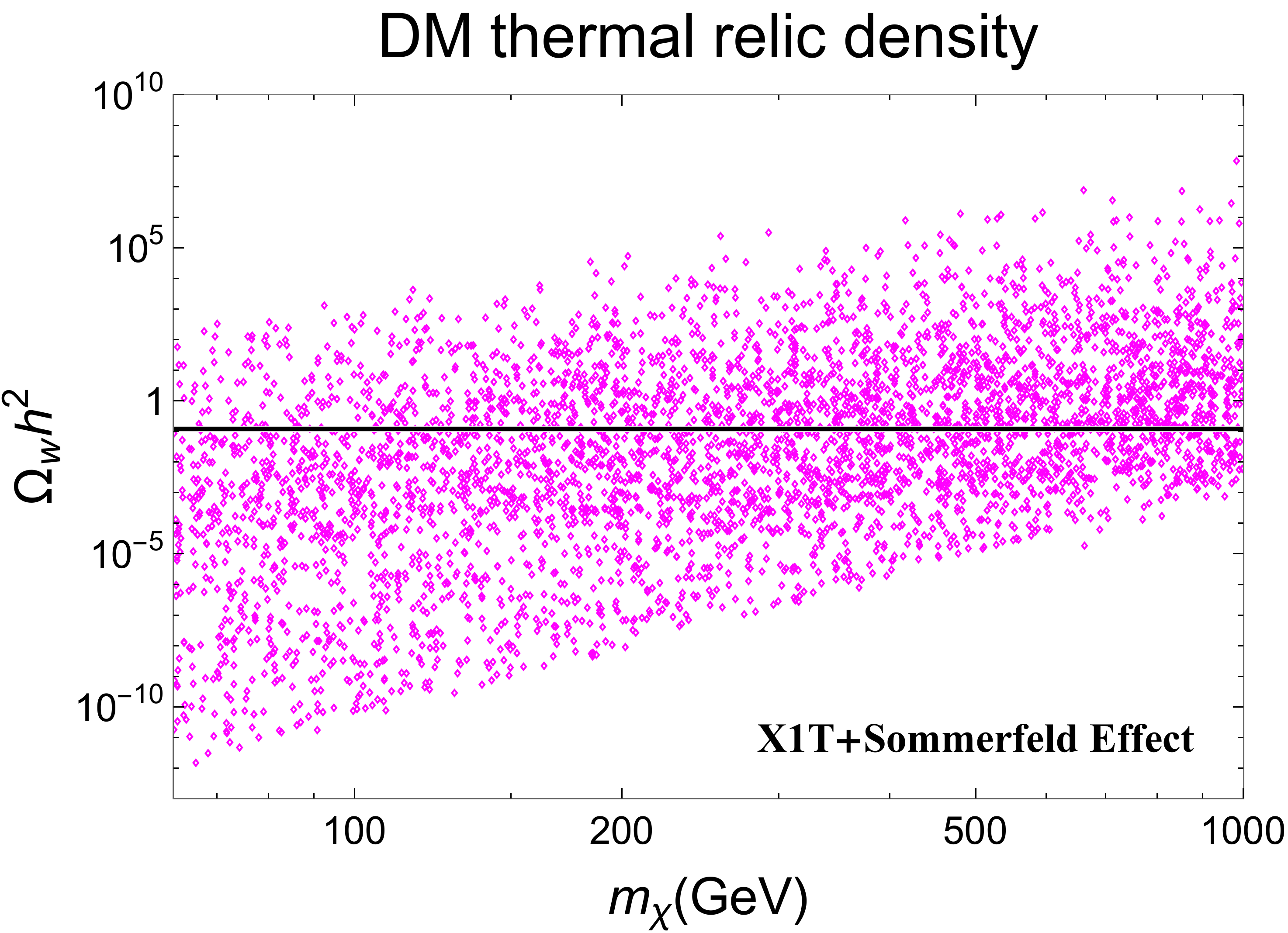} }
\caption{Relic density of leptonic scalar DM
    $\Omega_\chi h^2$ versus $m_\chi$: (a) without Sommerfeld
    enhancement, and (b) with Sommerfeld enhancement.}
\label{fig: Relic}
\end{figure}

Fig.~\ref{fig: Relic} shows the leptonic scalar DM thermal relic
density $\Omega_\chi h^2$ as a function of $m_\chi$: (a) without
Sommerfeld enhancement, and (b) with Sommerfeld enhancement effects.
We maintain the same symbols for each sample
used in Fig.~\ref{fig: direct} and Fig.~\ref{fig: Indirect}.
The horizontal line denotes the observed relic density:
$\Omega_{\rm obs}h^2 = 0.120$~\cite{Tanabashi:2018oca}.
Since the relic density is roughly proportional to the inverse of the total
$\langle\sigma_{\rm ann} v \rangle$, the samples 
are oriented reversely in vertical direction.
With Sommerfeld enhancement effect, the same parameters will lead to
a smaller relic density as expected. Thus, there are more regions of
the parameter space satisfying the relic density requirement
$\Omega_\chi h^2 < 0.123$~\cite{Tanabashi:2018oca}.

\section{Cosmological Constraints on the Right-handed Neutrino}

The left-handed neutrino decouple at the temperature $T_f^L\sim 1$ MeV 
when the left-handed neutrinos and the right-handed anti-neutrinos can  not been  converted to pairs of electron and positron.  
On the other hand, the right-handed neutrinos decouple at the temperature $T_f^R\sim m_\zeta$
when the production of the $\zeta$ particles is kinetically not allowed.  
At the temperature $T<T_f^L<T_f^R$, 
the total density of radiation $\rho_r$ is
\be
\rho_r=\rho_\gamma+\rho_L+\rho_R=3\Bigl [1+N_{\rm eff}\frac{7}{8}(\frac{4}{11})^{4/3} \Bigr ]\rho_\gamma,
\label{eq: Neff}
\en
where $\rho_\gamma$, $\rho_L$ and $\rho_R$ are the energy density of photons, $\nu_L$, and $\nu_R$, respectively. 
The relativistic degree of freedom $N_{\rm eff}$ here depends on the relativistic particle species 
and their internal degree of freedoms.  
Considering only three generations of left-handed neutrinos in the SM, 
the theoretical prediction is given by $N_{\rm eff}=3.045$~\cite{Mangano2005,Iocco2009}. 
The recent {\it Planck} 2018 data shows $N_{\rm eff}=2.92^{+0.36}_{-0.37}\ (95 \%{\rm CL})$ 
and this is compatible with the SM prediction.

Following the computation in Ref.~\cite{Zhang:2015wua},  
the additional contribution to the relativistic degree of freedom arising from $\nu_R$ 
is given by 
\be 
\Delta N_{\rm eff}=3 \times\left [\frac{g_{*s}(T_f^L)}{g_{*s}(T_f^R)}\right ]^{4/3}.
\label{eq:neff}
\en
Here, factor 3 describes three generations of neutrinos.  
As shown in Ref.~\cite{Zhang:2015wua}, 
the new relativistic degree of freedom is limited to be $\Delta N_{\rm eff}=0.10^{+0.44}_{-0.43}$.

We would like to note that both $\nu_L$ and $\nu_R$ completely decouple 
from the SM plasma before Big Bang nucleosynthesis (BBN) in this model,
thus one can obtain $g_{*s}(T_f^L)=10.75$ at $T_f^L\sim 1\mev$ and 
$g_{*s}(T_f^R)=67$ at $0.2\gev\lesssim T_f^R\lesssim 1.2\gev$.
By plugging these two values into Eq.~(\ref{eq:neff}), we can simply verify that 
$\Delta N_{\rm eff}=0.26$ also agrees with current limit~\cite{Tanabashi:2018oca}. 
Moreover, it has also pointed out in Ref.~\cite{Zhang:2015wua} that a combined constraint 
from Planck CMB data and BBN (the helium abundance measurements) 
reads $\Delta N_{\rm eff}\lesssim 0.53$ at $95\%$~C.L. which associates $T_f^R\gtrsim 200$~MeV. 
This implies $m_\zeta\gtrsim 200$~MeV if taking $T_f^R\simeq m_\zeta$. 
\textit{Therefore, to escape the combined constraint from CMB and BBN, 
we always take a safe limit $m_\zeta\gtrsim 200\mev$ in this work.}

\section{Small scale Problem}

As aforementioned in the introduction, 
the momentum transfer cross section $\sigma_T$ of the process $\chi\chi^*\to\chi^*\chi$ 
depends on the root mean square velocity ${v}_0$ of the DM particles. 
To solve the small scale (CCP/MSP/TBTF) problems, 
we simplify to use the following constraint~\cite{Rocha:2012jg,Elbert:2014bma}: 
\be
0.1~{\rm\ cm^2/g} \leq \langle\sigma_T/m_\chi\rangle_S \leq 10~{\rm\ cm^2/g},
\en
where $\langle\sigma_T/m_\chi\rangle_S$  is the Sommerfeld-enhanced cross section 
per unit DM mass. 

The process of $\chi\chi^*\to\chi^*\chi$ can occur via the the exchanges of $H$ or $\zeta$ in the $s$- as well as $t$-channels. It can also occur via the $\lambda_1$ quartic term interaction.
The subleading terms via the $s-$channel for the Sommerfeld effect can be ignored~\cite{Landau}. 
For the $t-$channel contribution, in the NR limit, we have 
\be
\sigma_T=\frac{(4\mu_{12}^2m_H^2+\lambda_{01}^2 {\rm v}^2 m_\zeta^2)^2}{64\pi m_\zeta^4 m_H^4 m_\chi^2}
\simeq\frac{\mu_{12}^4}{4\pi m_\zeta^4 m_\chi^2}.
\label{eq: sigmaT}
\en
The approximation holds only if $\mu_{12}\gg \lambda_{01}{\rm v}m_\zeta/({2m_H})$.  
We find that this approximation can be applied for the majority of our collected samplings. 
Note that we have ignored the quartic interaction 
since $\mu_{12}/m_\zeta\gg\lambda_{01}{\rm v} /({2m_H}) > \sqrt{\lambda_1/8}$. 
For example, we can have $\lambda_{01} {\rm v} /({2m_H})=1.85$ and
$\sqrt{\lambda_1/8}=0.67$
by taking $\lambda_{01}=\lambda_1=\sqrt{4\pi}$. 


\begin{figure}[t!]
\centering
\captionsetup{justification=raggedright}
\subfigure[]
{
  \includegraphics[width=0.4\textwidth,height=0.17\textheight]  {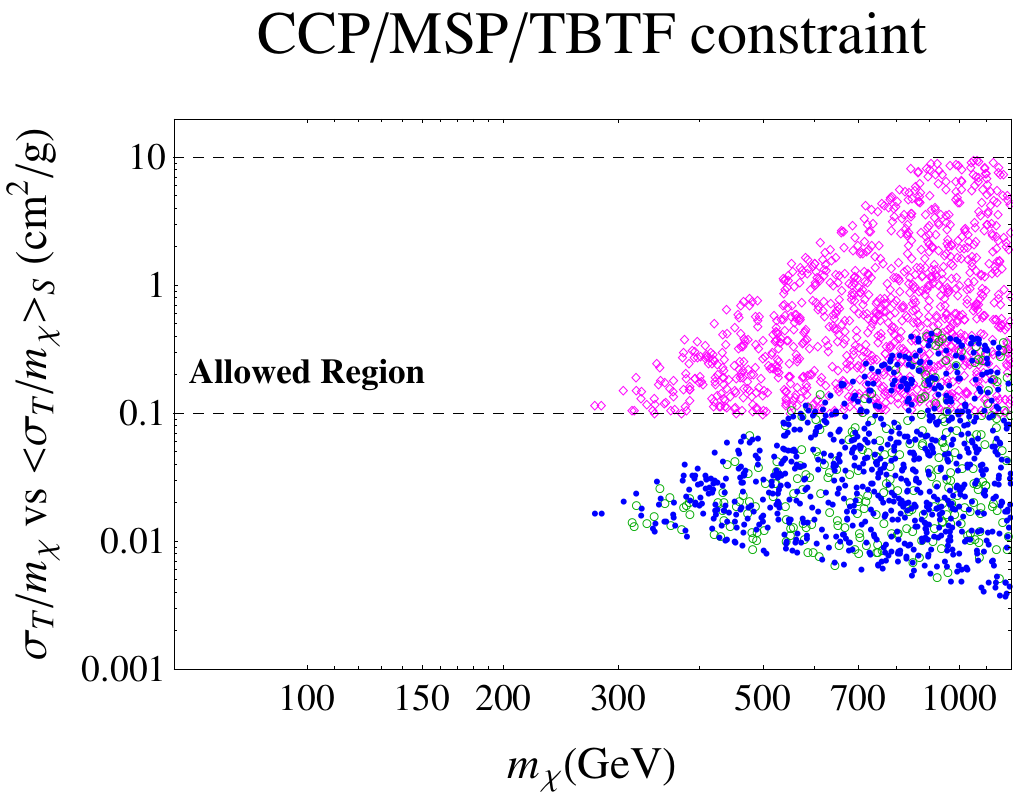}
}
\subfigure[]
{
  \includegraphics[width=0.4\textwidth,height=0.17\textheight]  {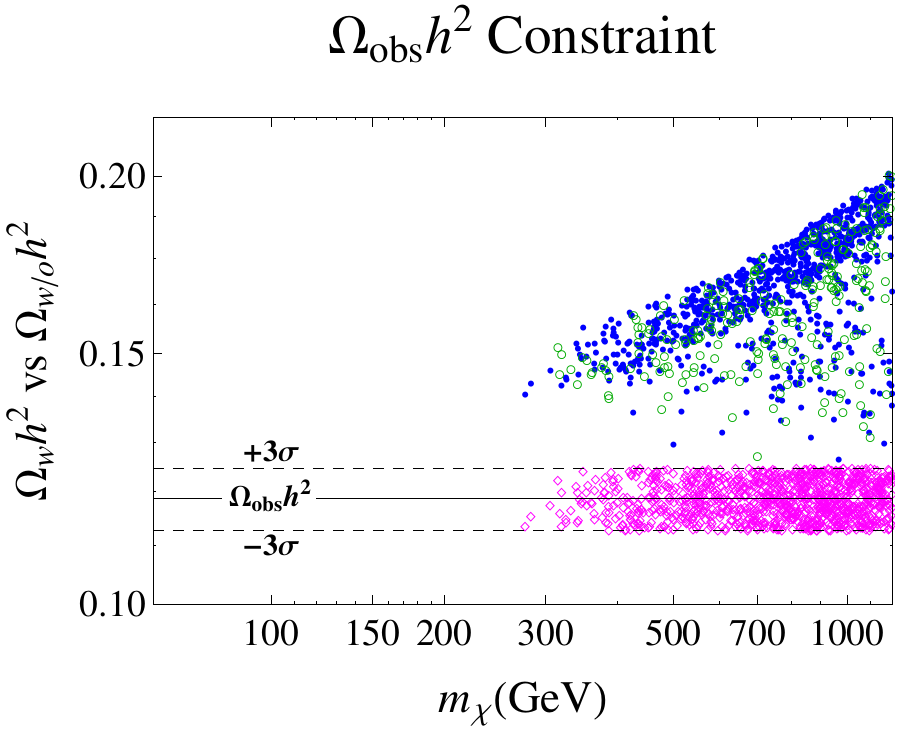}
}
\subfigure[]
{
  \includegraphics[width=0.4\textwidth,height=0.17\textheight]  {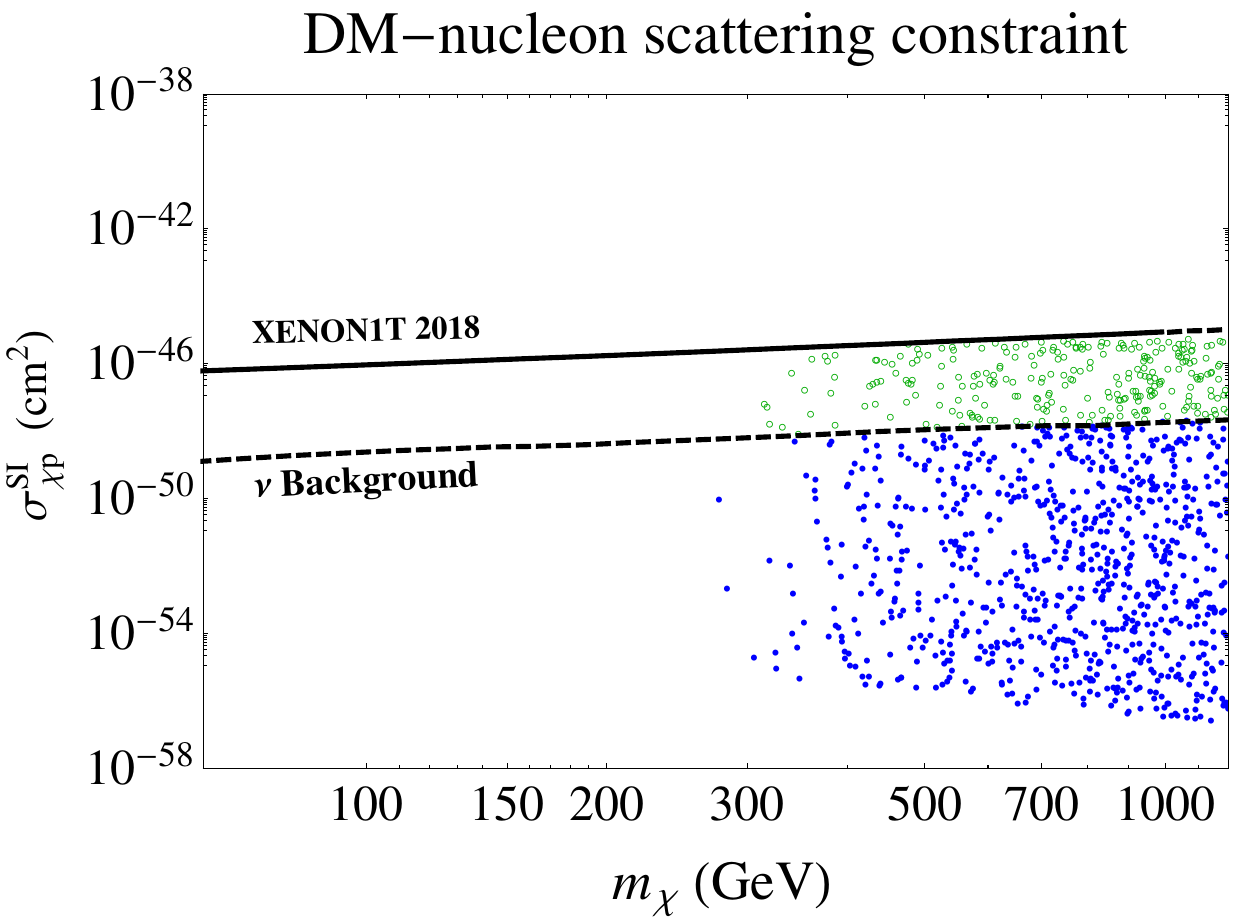}
}
\subfigure[]
{
  \includegraphics[width=0.4\textwidth,height=0.17\textheight]  {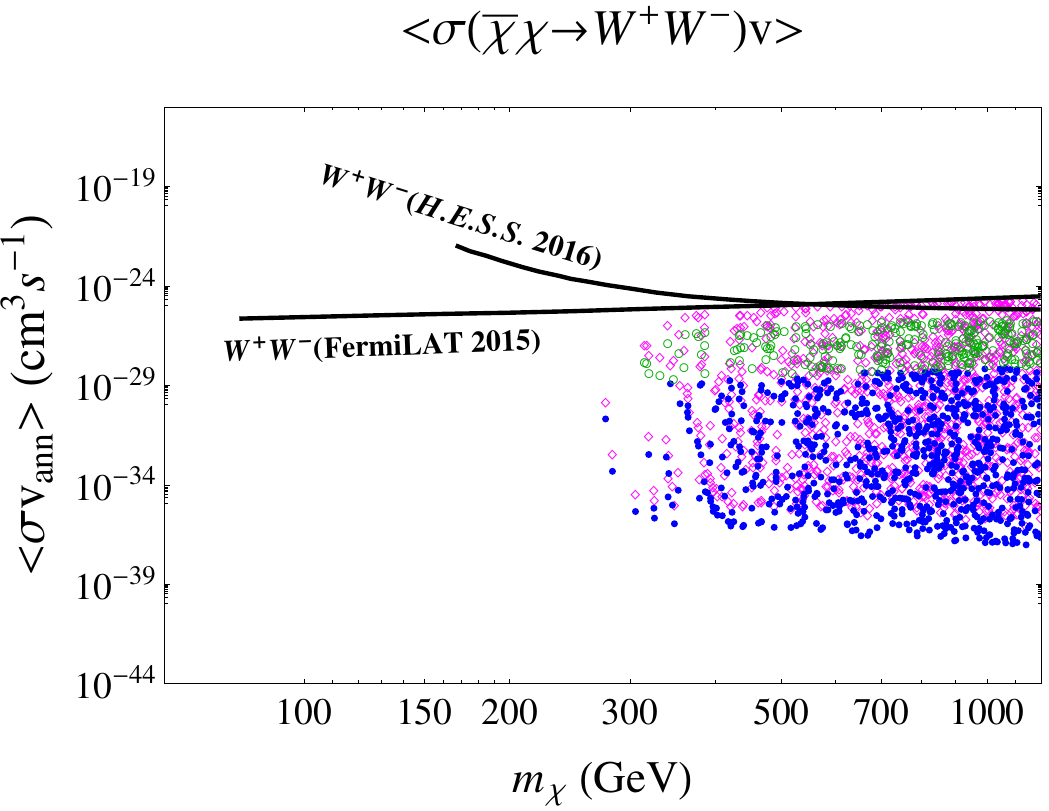}
}
\subfigure[]
{
 \includegraphics[width=0.4\textwidth,height=0.17\textheight]  {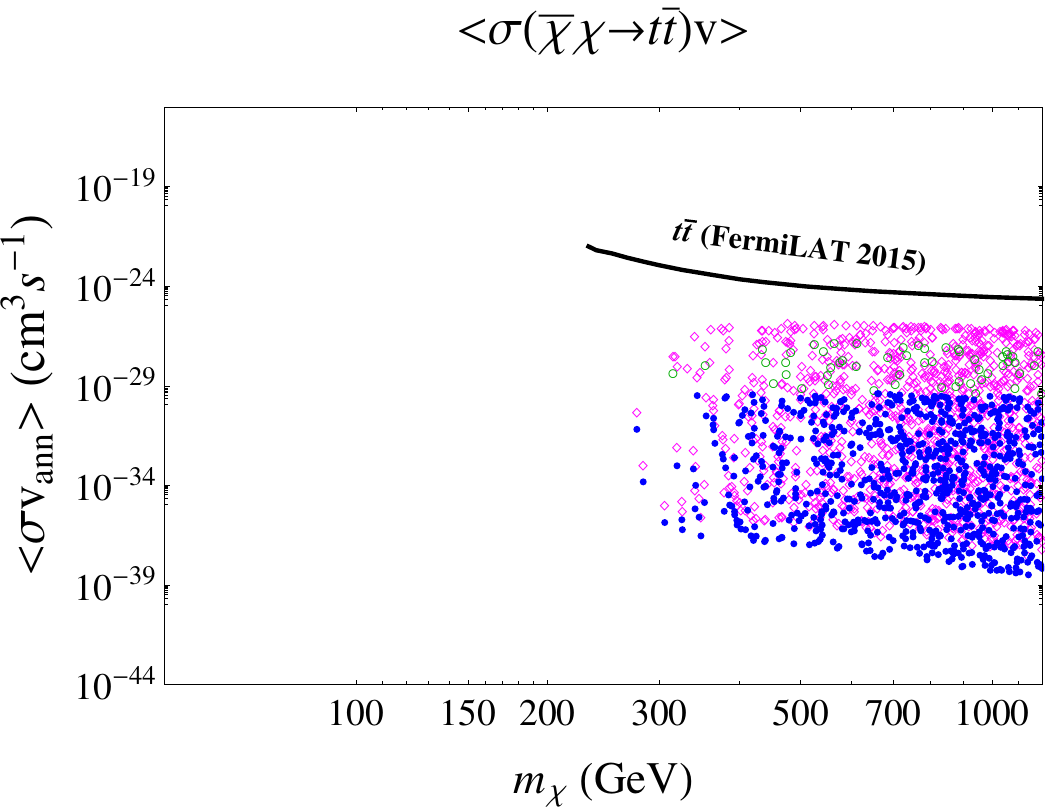}
}
\subfigure[]
{
 \includegraphics[width=0.4\textwidth,height=0.17\textheight]  {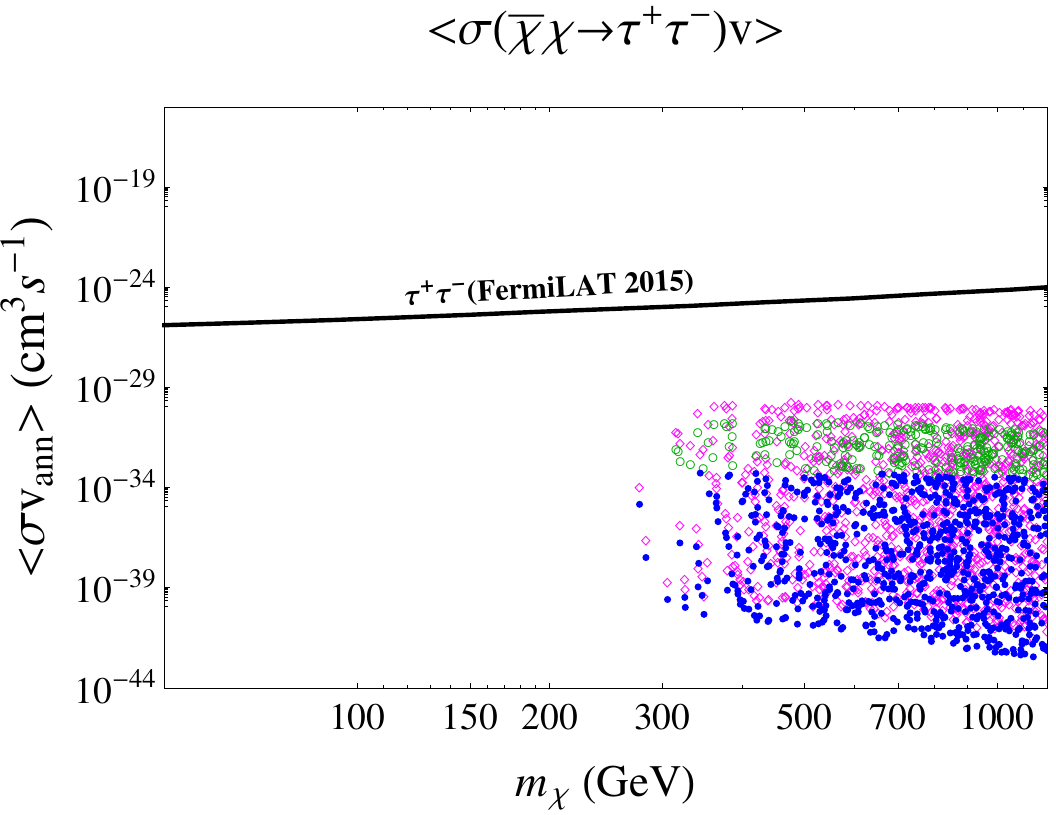}
}
\subfigure[]
{
 \includegraphics[width=0.4\textwidth,height=0.17\textheight]  {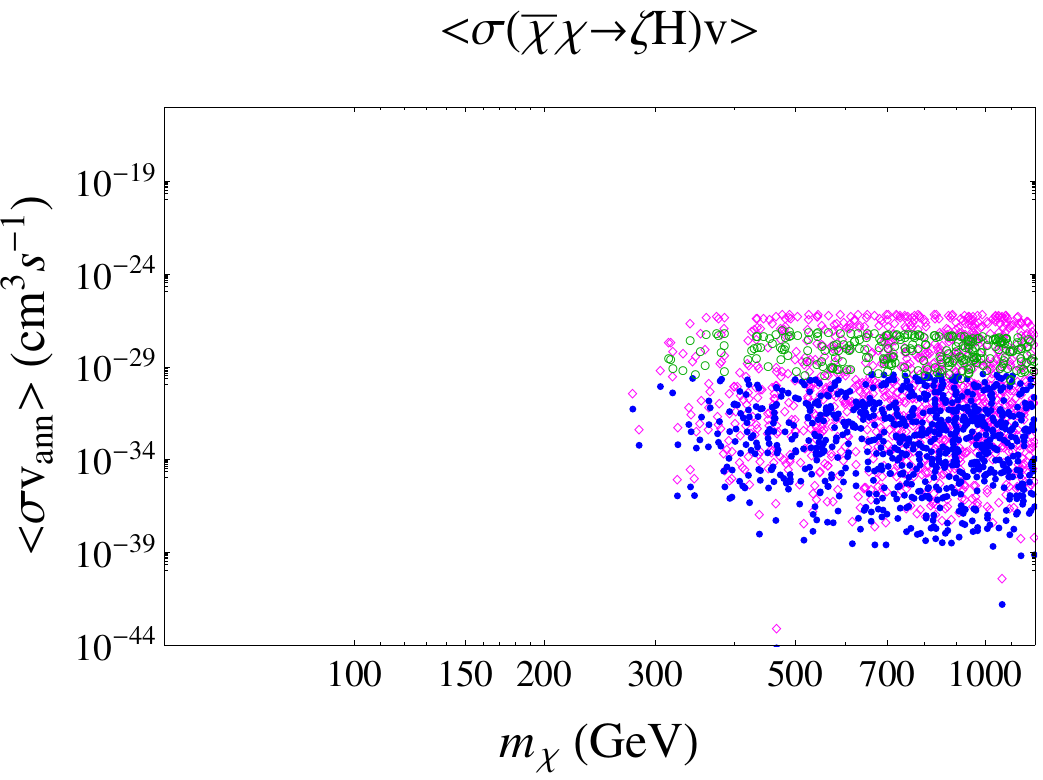}
}
\subfigure[]
{
 \includegraphics[width=0.4\textwidth,height=0.17\textheight]  {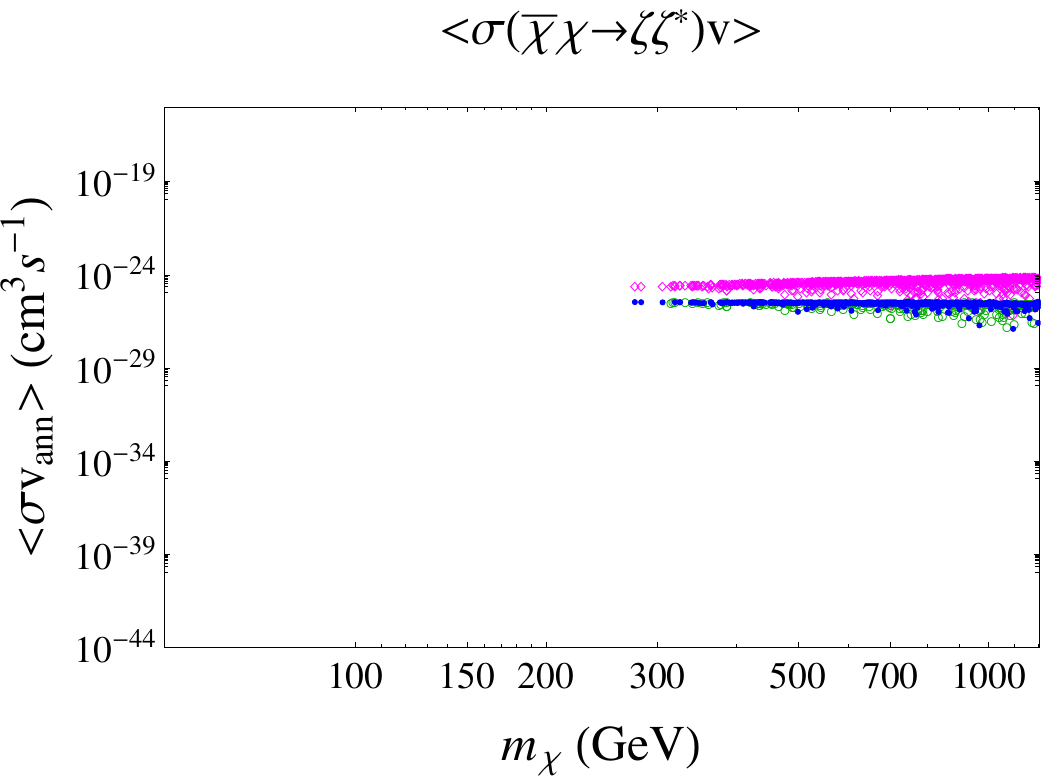}
}
\caption{Predictions on $\sigma_T/m_\chi$, $\Omega h^2$, $\sigma^{SI}$, and $\langle\sigma_{\rm ann} v\rangle$ in different channels for all allowed samples. ``${\color{green}\circ}$" and ``${\color{blue}\bullet}$" denote the samples which are testable for near future and below the neutrino floor, respectively, by the direct-detection experiment and without consideration of Sommerfeld effect.
``${\color{magenta}\diamond}$" denotes the values with considering the Sommerfeld effect}
\label{fig: renew}
\end{figure}

\section{Model Parameter Space}

In the CP-conserving LSDM model, there are eight free real parameters:
\be 
  m_\chi, m_\zeta, \mu_{12}, \lambda_{1}, \lambda_{2},
  \lambda_{01}, \lambda_{02}, \, {\rm and} \, \lambda_{12} \, .
  \nonumber
\en 
  In this section, we look for favored regions of the
parameter space in the LSDM model with implications from
astrophysical and cosmological observations.
We apply selection requirements for
small scale structure, cold dark matter relic density, direct searches,
and indirect detections, as well as cosmological constraints on right
handed neutrinos.

  For the leptonic scalar DM, the solution of small scale problem 
comes from DM strong self-interaction with $\sigma_T/m_\chi$
that contains 4 free parameters:
$m_\chi, m_\zeta, \mu_{12}$ and $\lambda_{01}$, 
as shown in Eq.~(\ref{eq: sigmaT}).
At first, we perform a grid scan for the ranges, 
$m_H/2<m_\chi<1200\gev$, $0.2<m_\zeta/\gev<1$, and
$-6<\log_{10}[\lambda_{01}]<\log_{10}[\sqrt{4\pi}]$.
With these collected samples, we can pin down the corresponding
allowed range for $\mu_{12}$ by using the allowed range of $\sigma_T/m_\chi$.
Second, we use the interpolation technique to find the
allowed range of $\mu_{12}$ from random sampling for $m_\chi, m_\zeta$ and
$\lambda_{12}$.
We then apply random sampling with the 8 free parameters as usual
except that $\mu_{12}$ is chosen from the allowed region
with randomly selected $m_\chi, m_\zeta$ and $\lambda_{12}$.
Third, we use the selected parameters to find the allowed samples
which satisfy the observed relic density constraint.
Finally, we find
the allowed parameter space by satisfying the constraints from the
direct and indirect searches.

Following this procedure, we collect a thousand samples that satisfy all the
constraints mentioned above.
Fig.~\ref{fig: renew} shows the predictions in the LSDM model for
$\sigma_T/m_\chi$, $\Omega_\chi h^2$, $\sigma^{SI}$, and
$\langle\sigma_{\rm ann} v\rangle$ in
$W^+W^-, t\bar t, \tau^+\tau^-, \zeta H$ and $\zeta\zeta^*$ channels.
Comparing the scenarios with and without Sommerfeld effect,
we only depict those samples in agreement with XENON1T data in Fig.~\ref{fig: renew}. 
The Sommerfeld effect is applied in the computation for 
the magenta samples ``${\color{magenta}\diamond}$", while 
Green ``${\color{green}\circ}$" (testable for near future) and
blue ``${\color{blue}\bullet}$" (below the neutrino floor) are 
obtained without including the Sommerfeld effect.
It is clear to see in Fig.~\ref{fig: renew}(a), that the
$\langle\sigma_T/m_\chi\rangle_S$ is enhanced by
the Sommerfeld effect such that all values fall into
$0.1\rm{\ (cm^2/g)} \leq
\langle\sigma_T/m_\chi\rangle_S \leq 10\rm{\ (cm^2/g)}$.
In Fig.~\ref{fig: renew}(b), we see that the relic density is roughly
proportional to $1/\langle\sigma_{\rm ann}\rangle$,
and hence the $\Omega_\chi h^2$ becomes suppressed
by the Sommerfeld effect
such that the relic density of all selected samples
fall into the range of $\Omega_{\rm obs} h^2 \pm 3\sigma$. 

It is interesting that the selected 1000 samples satisfying the
small scale requirement (CCP/MSP/TBTF) and the observed relic density
constraints also satisfy the the constraint on the SI DM-nucleon
scattering cross section $\sigma^{SI}_{\chi p}$ shown in Fig.~\ref{fig: renew}(c).
We have extended  XENON1T 2018 data to $m_{\chi} \simeq 1.2$ TeV
with the dashed line.
The cross section $\langle\sigma_{\rm ann} v\rangle$
for DM annihilating into
$W^+W^-$, $t\bar t$, $\tau^+\tau^-$, $\zeta H$ and $\zeta \zeta^*$
channels are presented in Fig.~\ref{fig: renew}(d)--Fig.~\ref{fig: renew}(h).
Combining indirect search in Fig.~(\ref{fig: Indirect})
and the small scale requirement, we see that the values of
$\langle\sigma_{\rm ann} v\rangle$ can not be too large, and
that $\chi\chi \to b\bar b$ and $\tau^+\tau^-$ may not be detectable.

Recall that the dominant $t$-channel cross section for
the self-interacting leptonic scalar DM at the tree level is 
\be 
\sigma(\chi\chi^* \to \chi\chi^* )
 = \frac{\mu_{12}^4}{4\pi m_\zeta^4 m_\chi^2}
 = \frac{g_\zeta^2 \mu_{12}^2}{4\pi m_\zeta^4 } \, .
\en
The dimensionless coupling $g_\zeta \equiv \mu_{12}/m_{\chi}$
is important to determine the annihilation cross section, and it
appears in the Yukawa potential [Eq.~(\ref{eq: Yukawa})]
contributing to the Sommerfeld enhancement effects.

\begin{figure}[t!]
\centering
\captionsetup{justification=raggedright}
{
 \includegraphics[width=0.48\textwidth,height=0.23\textheight]{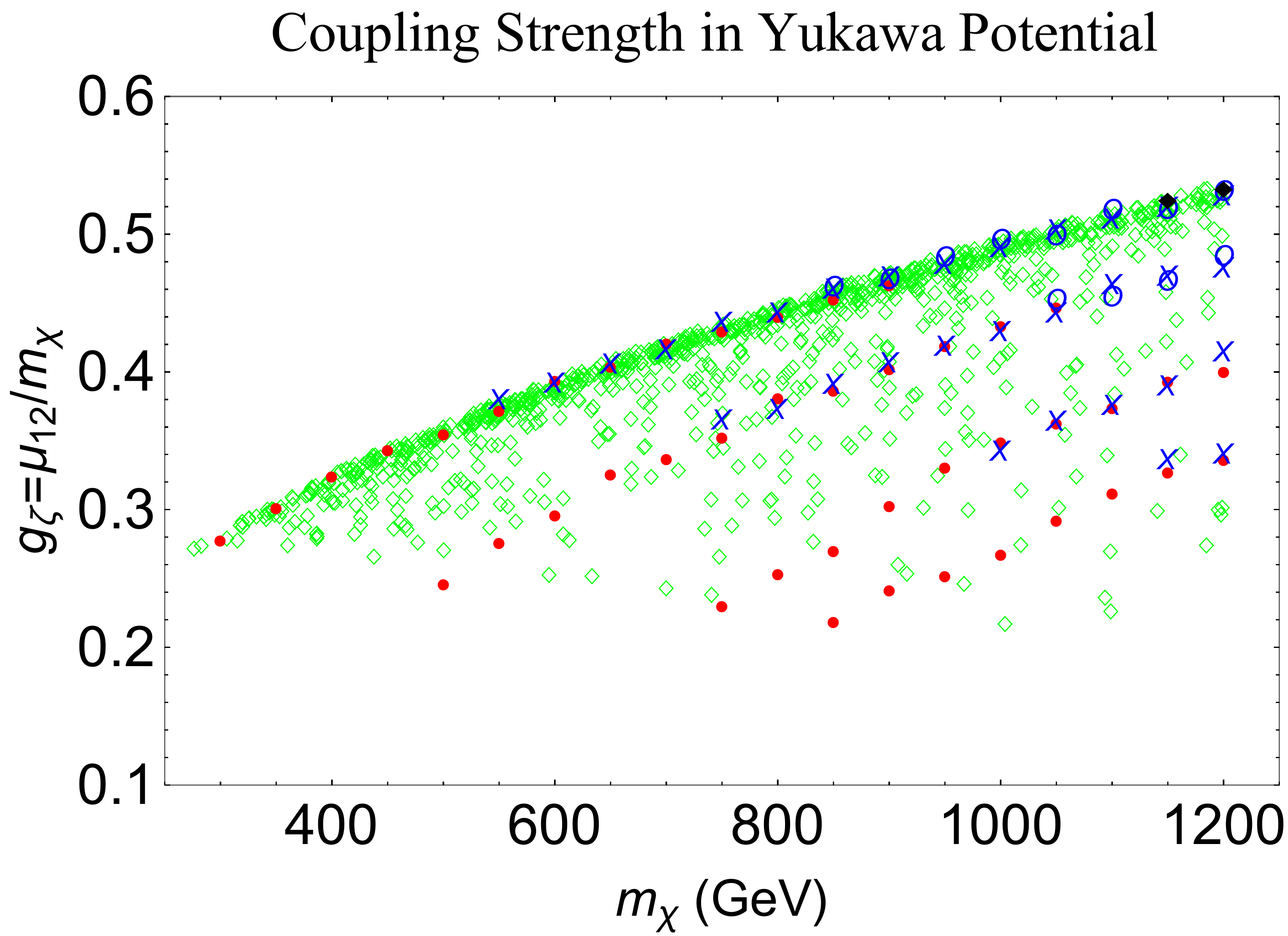}
}
\caption{The dimensionless coupling strength $g_\zeta \equiv
  \mu_{12}/m_{\chi}$ in the Yukawa potential versus DM mass $m_\chi$
  parametrized by the light mediator mass $m_\zeta$ where 
  $m_\zeta =$ 0.2 GeV (red ``${\color{red}\bullet}$"),
0.4 GeV (blue ``${\color{blue}\times}$"),
0.6 GeV (blue ``${\color{blue}\circ}$"), and
0.8 GeV (black ``${\color{black}\blacklozenge}$").}
\label{fig: all}
\end{figure}

The value of $g_\zeta$ is modified by Sommerfeld enhancement with
higher order and non-perturbative effects. 
Fig.~\ref{fig: all} presents $g_\zeta$ as a function of $m_\chi$
with several values of
$m_\zeta =$ 0.2 GeV (red ``${\color{red}\bullet}$"),
0.4 GeV (blue ``${\color{blue}\times}$"),
0.6 GeV (blue ``${\color{blue}\circ}$"), and
0.8 GeV (black ``${\color{black}\blacklozenge}$").
In addition, we consider four sets of parameters from top to bottom: 
$(\lambda_{01}, \lambda_{02}, \lambda_{12}, f_{RR})$=  $(10^{-6},
10^{-6},10^{-6},10^{-6})$, $(0.1,10^{-3}, 0.1, 0.0)$,  $(0.2, 10^{-3},
0.2, 0.2)$ and $(0.2,10^{-3},0.2,0.6)$, respectively. All allowed
samples consistent with all mentioned constraints are
denoted by green ``${\color{green}\diamond}$".
We find that $g_\zeta$ becomes larger with increasing value in
$m_\zeta$ or decreasing values in the parameter set
$(\lambda_{01}, \lambda_{02}, \lambda_{12}, f_{RR})$.
Note that for a given set of parameters, $\mu_{12}$ is
randomly sampled to satisfy all mentioned constraint.
The allowed range of $m_\zeta$ also depends on $m_\chi$.


From the 1000 allowed samples that are consistent with all mentioned
constraints, we can find the favored parameter space from the scanned
region.
BBN and CMB constraints require that 
$m_\zeta$ should be greater than 200 MeV, that leads to 
the minimal value of $m_\chi=276$ GeV.
The maximal value of $m_\zeta$ is found to be 814 MeV corresponding
to $m_\chi=1176$ GeV. 
$\lambda_1$ only involves the self-interaction process of
$\chi\chi*\to\chi\chi^*$ in Eq.~(\ref{eq: sigmaT}) and its contribution
can be ignored even with $\lambda_1=\sqrt{4\pi}$.
On the other hand,
$\lambda_2$ is irrelevant in our calculation.
We find that the maximal values of $\lambda_{01}$,  $\lambda_{02}$,
$\lambda_{12}$ and $f_{RR}$ are 0.27, 0.01 0.51 and 1.30,
respectively.
The allowed range of $\mu_{12}$ is between 75 and 634 GeV.

\section{Conclusions}

We have adopted a special model that has a leptonic scalar dark matter
(LSDM) ($\chi$) with lepton number $L_\chi = 1$ and a light scalar
mediator ($\zeta$) with $L_\zeta = 2$ and three flavors of neutrino
$\nu_R$ under the assumption of lepton number conservation.
In the early Universe,  DM thermalizes with
SM particles via $H$-portal and $\chi\chi^*\to\zeta\zeta^*$ provides
an efficient annihilation channel.
After DM freezes out, all $\zeta$ decay into $\nu_R\nu_R$
with a lifetime $\tau_\zeta\lesssim 10^{-11}$ (sec)
before the onset of BBN, and $\nu_R$ decouples from the SM particles
at the temperature $T_f^R \sim m_\zeta$.
The LSDM ($\chi$) and the light mediator ($\zeta$) with lepton number
conservation can provide a self-interacting  WIMP dark matter
that is consistent with astrophysical and cosmological constraints.

In the LSDM model, the $t-$channel exchange of a light mediator ($\zeta$)
makes the LSDM ($\chi$) self interacting cross section (SICS)
reasonable large. Furthermore, we evaluate the the Sommerfeld effects
and find significant enhancement for the SICS.
That makes the LSDM model suitable to explain the small scale structure of
the Universe.

We apply selection requirements for
small scale structure, cold dark matter relic density (Planck),
direct searches (XENON1T),
and indirect detections (Fermi-LAT and H.E.S.S),
as well as cosmological constraints on right-handed neutrinos.
A randomly selected set of parameters was found with 1000 samples
that satisfy all constraints. 
Large regions of the parameter space in the LSDM model 
are found to be consistent with astrophysical and cosmological
observations and collider Higgs properties.
A summary is in the following for the favored ranges of parameters: 
\begin{itemize}
\item 0.2 GeV $\lesssim m_\zeta \lesssim$ 0.814 GeV (BBN, CMB),
\item 276 GeV $\lesssim m_\chi \lesssim$ 1176 GeV (implied by $m_\zeta$),
\item 75 GeV $\leq \mu_{12} \leq$ 634 GeV, and 
\item $\lambda_{02} \le 10^{-2}$ (Higgs invisible width).
\end{itemize}
In addition, the upper bound of $\lambda_{01}$,  $\lambda_{02}$ and
$\lambda_{12}$ are 0.27, 0.01 and 0.51, respectively.

It is interesting that almost all regions of parameter space
satisfying astrophysical and cosmological observations lead to a
cold dark matter relic density with the most restrictive
requirement~\cite{Aghanim:2018eyx}.
\be
\Omega_{\rm c}h^2 \pm 3\sigma = 0.120 \pm 0.001
\en
that is
\be
0.117 \lesssim \Omega_\chi h^2 \lesssim 0.123.
\en
A more realistic requirement should be
\be
\Omega_\chi h^2 \lesssim 0.123 \, .
\en
That will enlarge the favored parameter space and accommodate more
types of dark matter particles.

\section*{Acknowledgments}

We are grateful to Ernest Ma for beneficial discussion.
This research was supported by
grants MOST 109-2811-M-002-530 and NTU 108L104019 (GGW),
as well as in part by the U.S. Department of Energy and the University of
Oklahoma (CK). 
Y.-L.~S.~Tsai was funded by the Ministry of Science and Technology Taiwan under Grant
No.~109-2112-M-007-022-MY3.

\appendix

\section{Sommerfeld enhancement in $\chi\chi$ annihilations}
\label{sec:SE}

\subsection{Bethe-Salpeter equation in $\chi\chi\to \chi\chi$ process}
\label{sec:Bethe_Salpeter}
The Feynman diagram of non-perturbative scattering $\chi(p_1)\chi(p_2)\to \chi(p_3')\chi(p_4')$ is shown in Fig.~\ref{fig: Som2}(a).
Note that $p'_3$ and $p'_4$ are not necessary on-shell as these two lines will be connected to $\chi\chi$ annihilation diagrams later [see Fig.~\ref{fig: Som2}(b)].
Following the standard procedure~\cite{Landau,ChuaWong2017}, we will derive the Bathe-Salpeter equation for the process of scalar DM scattering via the scalar $H$-exchange repeatedly.

The amputated non-perturbative 4-point vertex function can be written as
\be
\hspace{-0.5cm}
i\Gamma(p'_3, p'_4; p_1, p_2) 
&=&i\tilde\Gamma(p'_3, p'_4; p_1, p_2)\non\\
&&+\int \frac{d^4p''_3}{(2\pi)^4}
i\tilde\Gamma(p'_3,p'_4;p''_3,p''_4)
(iD_{F}(p''_3)) 
i\Gamma(p''_3,p''_4;p_1,p_2)(iD_F(p''_4)),
\label{eq:master0}
\en
where we have $p''_4=p'_3+p'_4-p''_3$, $D_{F}$ is the scalar DM propagator
and, the amputated tree-level 4-point vertex function through the $H$-exchange is given by
\be
i\tilde\Gamma(p'_3,p'_4;p''_3,p''_4)=-i \lambda_{01}^2 v^2\frac{1}{(p''_3-p'_3)^2-m^2_H}.
\en 
With the instantaneous approximation,
namely, ignoring the time component of the momentum transfer, the tree-level 4-point vertex function is just the potential $U_H(\vec p\ ''_3-\vec p\ '_3)$ defined below
\be
i\tilde\Gamma(p'_3,p'_4;p''_3,p''_4)
= i \lambda_{01}^2 v^2\frac{1}
{(\vec p\ ''_3-\vec p\ '_3)^2+m^2_H}\equiv i U_H(\vec p\ ''_3-\vec p\ '_3 ).
\label{eq:vertex}
\en
To proceed we define two auxiliary functions as follows~\cite{ChuaWong2017}:
\be
i\eta(p_3,p_4;p_1,p_2)
&\equiv&iD_{F}(p_3) i\Gamma(p_3,p_4;p_1,p_2)(iD_{F}(p_4)),
\non\\
i\tilde\chi(p_3,p_4;p_1,p_2)
&\equiv&iD_{F}(p_3) i\tilde\Gamma(p_3,p_4;p_1,p_1)(iD_{F}(p_4)),
\label{eq:vertex2}
\en
and Eq.~(\ref{eq:master0}) can be expressed as
\be
i\eta(p'_3,p'_4;p_1,p_2)
&=&
i\tilde \chi(p'_3,p'_4;p_1,p_2)
\non\\
&&+
\int \frac{d^4p''_3}{(2\pi)^4}
i\tilde\chi(p'_3,p'_4;p''_3,p''_4)
i\eta(p''_3,p''_4;p_1,p_2).
\label{eq: ieta}
\en
Adding  $(2\pi)^4\delta^4(p''_3-p_1)$
to both side of the above equation,
and defining 
\be
i\chi(p'_3,p'_4;p_1,p_2)&\equiv&
 (2\pi)^4\delta^4(p''_3-p'_3)+i\eta(p'_3,p'_4;p_1,p_2),
\label{eq: ichidefine1}
\en
Eq. (\ref{eq: ieta}) becomes
\be
i\chi(p'_3,p'_4;p_1,p_2)
=
(2\pi)^4\delta^4(p''_3-p'_3)
+
\int \frac{d^4p''_3}{(2\pi)^4}
i\tilde\chi(p'_3,p'_4;p''_3,p''_4)
i\chi(p''_3,p''_4;p_1,p_2).
\label{eq: ichi11}
\en

In the NR limit, the scalar propagator can be approximately written as
\be
D_F(\pm k)=\frac{1}{2m_\chi}\frac{1}{(k_0-m_\chi)-\vec k^2/2m_\chi+i\epsilon}\equiv \frac{1}{2m_\chi}g_\chi(k),
\en
Substitute it into the above equation, we obtain the equation for $\chi$, 
\be
i\chi(p'_3,p'_4;p_1,p_2)
=(2\pi)^4\delta^4(p''_3-p'_3)
+\frac{1}{4m_\chi^2}g_\chi(p'_3)g_\chi(p'_4)\int \frac{d^4p''_3}{(2\pi)^4}
U_H(\vec p''_3-\vec p'_3)
\chi(p''_3, p''_4 ;p_1,p_2).\non\\
\label{eq: ichi31}
\en
In fact, we can drop the redundant variables $p_1$ and $p_2$ in the above equation.
Now we define
\be 
\left\{ \begin{array}{rcl}
&&p\equiv (p'_3-p'_4)/2\\
&&P\equiv (p'_3+p'_4)/2\\
&&\hat\chi(k_1, k_2)\equiv \chi(k_1+k_2, k_1-k_2)\\
\end{array}\right..
\en
Eq.~(\ref{eq: ichi31}) can be rewritten as 
\be
i\hat\chi(P, p)=(2\pi)^4\delta^4(q-P-p)
+\frac{1}{4m_\chi^2}g_\chi(P+p)g_\chi(P-p) \int\frac{d^4q}{(2\pi)^4}U_H(\vec q-\vec P-\vec p)\hat\chi(P,q-P).\non\\
\en
Let $q'=q-P$ and redefine $q'=q$. The above equation becomes
\be
i\hat\chi(P, p)=(2\pi^4)\delta^4(q-p)
+\frac{1}{4m_\chi^2}g_\chi(P+p)g_\chi(P-p) \int\frac{d^4q}{(2\pi)^4}U_H(\vec q-\vec p)\hat\chi(P,q).
\label{eq: chihateq}
\en
Defining the Bathe-Salpeter wave function as
\be
\psi(\vec q)\equiv\int\frac{dq_0}{2\pi}i\hat\chi(P, q),
\en
and integrating with respect to $p_0$ on both sides of Eq.~(\ref{eq: chihateq}),
we have
\be
\psi(\vec p)=(2\pi)^3\delta(\vec q-\vec p)
+\frac{1}{2\pi i}\frac{1}{4m_\chi^2}\int dp_0\ g_\chi(P+p)g_\chi(P-p)\int\frac{d^3\vec q}{(2\pi)^3}U_H(\vec q-\vec p)\psi(\vec q).\non\\
\label{eq: psieq}
\en

By taking $p'_3=(E'_3, \vec p)$, $p'_4=(E'_4,-\vec p)$, $p''_3=(E''_3, \vec p)$ and $p'_4=(E''_4,-\vec p)$ in the center of mass frame, we have $p\equiv (\epsilon, \vec p)
=((E'_3-E'_4)/2, \vec p)$ and $P\simeq(m_\chi+E/2, \vec 0)$ and the total kinetic energy $E=\mu v^2/2$ where $\mu=m_\chi/2$ is the reduce mass of $\chi\chi$ system and the relative velocity $v=v_{\rm lab}$ defined in Eq.~\ref{eq:vlab}.
Using the residue theorem, we integrate over $p_0=\epsilon$ in Eq.(\ref{eq: psieq}), and obtain 
\be
\psi(\vec p)=(2\pi)^3\delta(\vec q-\vec p)
+\frac{1}{E-\frac{\vec p^2}{m_\chi}}
\int\frac{d^3\vec q}{(2\pi)^3}V_H(\vec q-\vec p)\psi(\vec q),\quad V_H(\vec q)\equiv -\frac{1}{4m_\chi^2}U_H(\vec q),
\en
The above equation is simply the Bathe-Salpeter equation in the momentum space representation.
By taking the Fourier transformation, we have
\be
\int\frac{d^3\vec p}{(2\pi)^3}e^{-i\vec p\cdot\vec r}
(\frac{\vec p^2}{m_\chi}-E)
[\psi(\vec p)+(2\pi)^3\delta(\vec q-\vec p)]
+
\int\frac{d^3\vec p}{(2\pi)^3}e^{-i\vec p\cdot\vec r}
\int\frac{d^3\vec q}{(2\pi)^3}V_H(\vec q-\vec p)\psi(\vec q)=0.\non\\
\en
After simplification, we obtain the Bathe-Salpeter equation in the position representation:
\be
-\frac{1}{2\mu}\nabla^2\psi(\vec r)+V_H(r)\psi(\vec r)=E\psi(\vec r),\quad V_H(r)=-\alpha_H\frac{e^{-m_H r}}{r},
\label{eq: BS}
\en
where we see the potential is Yukawa-type with the corresponding fine structure constant
$\alpha_H=g_H^2/4\pi$  and 
the dimensionless coupling strength $g_H=\lambda_{01} v/2 m_\chi$.

\subsection{s-wave Sommerfeld factor in $\chi\chi\to \nu_R\nu_R$ process}

From Fig.~\ref{fig: Som2}(b), the Sommerfeld enhanced amplitude $iA_{S}$ of $\chi\chi\to \nu_R\nu_R$ annihilation process can be expressed as 
\be
iA_{S}(p_3, p_4, p_1, p_2)
&=&iA(p_3, p_4, p_1, p_2) \non\\
&+&\int \frac{d^4p'_3}{(2\pi)^4}
iA(p_3,p_4;p'_3,p'_4)
(iD_{F}(p'_3))
i\Gamma(p'_3,p'_4;p_1,p_2)
(iD_{F}(p'_4)) 
\label{eq:wholeM2}
\en
where 
$iA$ is the amplitude of the process at tree level. 
With the help of Eq.~(\ref{eq:vertex2}), we have
\be
iA_{S}(p_3, p_4; p_1, p_2)
=
\int \frac{d^4p'_3}{(2\pi)^4}
iA(p_3,p_4;p'_3,p'_4)
i\chi(p'_3,p'_4;p_1,p_2).
\label{eq:whole}
\en
For the s-wave rescattering, the amplitude $A(p_3,p_4;p'_3,p'_4)$ is independent of momentum and hence, the above equation becomes
\be
iA_{S}(p_3, p_4,;p_1, p_2)
&=&
iA(p_3,p_4;p_1,p_2)
\int \frac{d^3\vec p}{(2\pi)^4}
\int\frac{dp_0}{(2\pi)}i\hat\chi(P,p)
\non\\
&=&
iA(p_3,p_4;p_1,p_2)\psi_{l=0}(\vec r=0)
\label{eq:whole2}
\en
Hence we have
\be
|A_{S}(p_3, p_4,;p_1, p_2)|^2=|A_(p_3, p_4,;p_1, p_2)|^2S,
\en
where $S=|\psi_{l=0}(\vec r=0)|^2$ is the so-called s-wave Sommerfeld factor and the wave function 
$\psi(\vec r)$ satisfies the Bathe-Salpeter Equation in Eq.~(\ref{eq: BS}).

\subsection{Solving $\psi(\vec r=0)$ numerically}
\label{sec:numS}
Let us consider the general case. As we know that two DM particles form a bound $\chi\chi$ or $\chi\chi^*$ state before annihilation. 
This two-particle wave function $\psi(\vec r)$ satisfies the following Schr\"odinger equation [see in Eq.~(\ref{eq: Schrodinger eq}) ]
\be
-\frac{1}{2\mu} \nabla^2\psi(\vec r)+V(\vec r)\psi(\vec r)=E\psi(\vec r)=\frac{1}{2} \mu v^2\psi(\vec r).
\label{eq: Schrodinger2}
\en
The separation of variables give us the radial Schr\"odinger equation
\be 
[-\frac{1}{2\mu}\frac{1}{r^2}\frac{d}{dr}(r^2\frac{d}{dr})+V(r)+\frac{l(l+1)}{2\mu r^2}]R_l(r)=ER_l(r).
\label{eq: RadialA}
\en
Here we follow~\cite{Iengo:2009ni, ChuaWong2017} to solve for $\psi(\vec r=0)$ numerically. 
From the scattering theory, the radial wave function has the following asymptotic form~\cite{Sakurai}:
\be
R_l(r)\rightarrow 
e^{i\delta_l}
\frac{\sin(pr-l\pi/2+\delta_l)}{pr},
\label{eq: asymptotic R}
\en
where $\delta_l$ is the phase shift corresponding to the partial wave with angular momentum quantum number $l$.
Defining $\Phi_l$ by $R_l(\rho)=N \rho^l \Phi_l(r)$ with $\rho=p r$, and normalization constant $N$ to be determined later, 
Eq.~(\ref{eq: RadialA}) becomes
\be
\Phi^{\prime\prime}_l+\frac{2(l+1)}{\rho}\Phi'_l+(-\frac{2}{p v}V(r)+1)\Phi_l=0,
\en
where the initial conditions are taken to be~\cite{Iengo:2009ni}
\be
\Phi_l(0)=1, 
\quad
\Phi'_l(0)=\frac{\rho V(r)}{pv (l+1)}\bigg|_{\rho\to {0}}\Phi_l(0),
\en
for a regular solution.
We now concentrate on the $l=0$ case. 
As one can see by taking $\rho\gg 1$, 
in the case that $|\rho V(r)|\ll1$,
the differential equation and its solution become
\be
\Phi^{\prime\prime}_0+\frac{2}{\rho}\Phi'_0+\Phi_0\bigg|_{\rho\gg 1}=0,
\qquad
\Phi_0(\rho)\to C\frac{\sin(\rho+\delta_0)}{\rho},
\en
with $C$ a real number. 
The above $\Phi_0$ is to be compared to $R_0(r)\to e^{i\delta_0}\sin(\rho+\delta_0)/\rho$
[see Eq. (\ref{eq: asymptotic R})], as $\rho\gg 1$.
To work out the normalization $N$, it is useful noting, in the $\rho\gg 1$ region,
\be
\Phi_0(\rho-\pi/2)\to -C\frac{\cos(\rho+\delta_0)}{\rho-\pi/2},
\en
which can be used with $\Phi_0(\rho)$ to construct 
\be
\kappa\equiv\lim_{\rho\to\infty}e^{i\rho}\left[{-i\rho\Phi_0(\rho)-(\rho-\pi/2)\Phi_0(\rho-\pi/2)}\right]
=C e^{-i\delta_0}.
\label{eq: kappa}
\en
Consequently, we see that $R_0(r)$ can be obtained as
\be
R_0(r)=\kappa^{-1}\Phi_0(\rho),
\en 
since it satisfies the Schr\"{o}dinger equation and has the correct asymptotic behavior.
Finally, we have
\be
\psi(\vec r=0)=\kappa^{-1}\Phi_0(0)=\kappa^{-1}=\lim_{\rho\to\infty}\frac{e^{-i\rho}}{-i\rho\Phi_0(\rho)-(\rho-\pi/2)\Phi_0(\rho-\pi/2)}.
\label{eq: psikappa}
\en
Note that the phase of $\psi(\vec r=0)$ is just $\delta_0$ [see Eq. (\ref{eq: kappa})].

Now we are ready to do the numerical calculation. For a scalar DM $\chi$ with a scalar mediator $X$, we have the Yukawa-type potential:
\be
V_X(r)=-\alpha_{X}\frac{e^{-m_X r}}{r},
\label{eq: YukawaX}
\en
where $\alpha_X=g_X^{'2}/(16\pi m_\chi^2)$ with the dimensionful coupling strength $g'_X=\lambda_{01}{\rm v}$, namely, the dimensionless coupling strength $g_X=g'_X/2m_\chi$. Hence we need to solve the following differential equation:
\be
\Phi''(\rho)+\frac{2}{x}\Phi'(\rho)+(1-\frac{2 a_X e^{-b_X \rho}}{\rho})\Phi(\rho)=0,
\en
with the boundary conditions:
\be 
\Phi(\rho=0)=1,\quad \Phi'(\rho=0)=-a_X.
\en
In the above $a_X=\alpha_X$ and $b_X=2 m_X/m_\chi v$.
We find that it is enough to take $\rho\simeq 200$ to obtain the limit in Eq. (\ref{eq: psikappa}).

\end{document}